\pdfoutput=1
\documentclass[journal ]{new-aiaa}
\usepackage[utf8]{inputenc}

\usepackage{textcomp}

\usepackage{graphicx}
\usepackage{amsmath}
\usepackage[version=4]{mhchem}
\usepackage{siunitx}
\usepackage{longtable,tabularx}
\usepackage{multicol}
\usepackage{caption}
\usepackage{placeins}

\hypersetup{
  colorlinks= true,
  linkcolor = blue,
  urlcolor  = blue,
  citecolor = black,
}
\newcommand*{\fullref}[1]{\hyperref[{#1}]{\autoref*{#1} \nameref*{#1}}}
\newcommand*{\secref}[1]{\hyperref[{#1}]{\autoref*{#1}}}

\graphicspath{{images/}}
\usepackage{wrapfig}
\usepackage{enumitem}
\newlist{inlineenum}{enumerate*}{1}
\setlist*[inlineenum]{mode=unboxed,label=(\arabic*)}

\newlist{inlineitem}{itemize*}{1}
\setlist*[inlineitem]{label=\textbullet}

\usepackage{multirow,tabularx}
\newcolumntype{Y}{>{\centering\arraybackslash}X}
\usepackage{multirow, makecell}
\usepackage{booktabs} 

\usepackage{placeins}
\usepackage[all]{nowidow}
\usepackage{multicol}
\usepackage{listings}

\newcommand{\wrapfill}{\par\ifnum\value{WF@wrappedlines}>0
  \addtocounter{WF@wrappedlines}{-1}%
  \null\vspace{\arabic{WF@wrappedlines}\baselineskip}%
  \WFclear
\fi}

\usepackage[acronym]{glossaries} 

\usepackage{glossary-mcols}
\newcommand*\myglsentry[1]{%
  \protect\ifglsused{#1}{%
    \glsentryshort{#1}%
  }{%
    \glsentrylong{#1}%
  }%
}
\newcommand*\myglsentryplural[1]{%
  \protect\ifglsused{#1}{%
    \glsentryplural{#1}%
  }{%
    \glsentrylongpl{#1}%
  }%
}

\makeglossaries

\newacronym{3d}{3D}{three-dimensional}
\newacronym{aiaa}{AIAA}{American Institute of Aeronautics and Astronautics}
\newacronym{ai}{AI}{Artificial Intelligence}
\newacronym{anfis}{ANFIS}{Adaptive Neuro-Fuzzy Inference System}
\newacronym{atmonto}{ATMONTO}{Air Traffic Management Ontology}
\newacronym{bfo}{BFO}{Basic Formal Ontology}
\newacronym{cad}{CAD}{Computer-Aided Design}
\newacronym{catia}{CATIA}{Computer Aided Three Dimensional Interactive Application}
\newacronym{clif}{CLIF}{Common Logic Interchange Format}
\newacronym{cmdows}{CMDOWS}{Common \myglsentry{mdo} workflow schema}
\newacronym{cpacs}{CPACS}{Common Parametric Aircraft Configuration Schema}
\newacronym{csv}{CSV}{Comma-separated values}
\newacronym{dl}{DL}{Description Logic}
\newacronym{dlr}{DLR}{Deutsches Zentrum für Luft- und Raumfahrt}
\newacronym{dtad}{DTAD}{Digital Thread for Airplane Design}
\newacronym{fair}{FAIR}{Findable, Accessible, Iinteroperable, Reusable}
\newacronym{json}{JSON}{JavaScript Object Notation}
\newacronym{json-ld}{JSON-LD}{JSON for Linked Data}
\newacronym{kadmos}{KADMOS}{Knowledge- and graph-based Agile Design with \myglsentry{mdo} System}
\newacronym{kbe}{KBE}{Knowledge-Based Engineering}
\newacronym{kif}{KIF}{Knowledge Interchange Format}
\newacronym{km}{KM}{knowledge management}
\newacronym{mdao}{MDAO}{multi-disciplinary analysis and optimization}
\newacronym{mdax}{MDAx}{MDAO Workflow Design Accelerator}
\newacronym{mdo}{MDO}{multi-disciplinary optimization}
\newacronym{mbse}{MBSE}{model-based systems engineering}
\newacronym{nasa}{NASA}{National Aeronautics and Space Administration}
\newacronym{omg}{OMG}{Object Management Group}
\newacronym{orkg}{ORKG}{Open Research Knowledge Graph}
\newacronym{owl}{OWL}{Web Ontology Language}
\newacronym{pido}{PIDO}{Process Integration and Design Optimization}
\newacronym{plm}{PLM}{Product Lifecycle Management}
\newacronym{radex}{RaDEX}{Rationale-based Design Explanation}
\newacronym{rce}{RCE}{remote component environment}
\newacronym{rdf}{RDF}{Resource Description Framework}
\newacronym{rdfa}{RDFa}{RDF in Attributes}
\newacronym{se2a}{SE²A}{Sustainable and Energy-Efficient Aviation}
\newacronym{slr}{SLR}{Systematic Literature Review}
\newacronym{sparql}{SPARQL}{SPARQL Protocol And \myglsentry{rdf} Query Language}
\newacronym{step}{STEP}{Standard for the Exchange of Product model data}
\newacronym{swarm-slr}{SWARM-SLR}{ Streamlined Workflow Automation for Machine-actionable \myglsentryplural{slr}}
\newacronym{swrl}{SWRL}{Semantic Web Rule Language}
\newacronym{sysml}{SysML}{Systems Modeling Language}
\newacronym{tu}{TU}{Technische Universität}
\newacronym{uml}{UML}{Unified Modeling Language}
\newacronym{uri}{URI}{Uniform Resource Identifier}
\newacronym{w3c}{W3C}{World Wide Web Consortium}
\newacronym{xml}{XML}{Extensible Markup Language}

\title{Knowledge-Based Aerospace Engineering - \\A Systematic Literature Review}

\author{Tim Wittenborg\footnote{PhD Student, L3S Research Center, Appelstr. 9a, 30167 Hannover, tim.wittenborg@l3s.de} and Ildar Baimuratov\footnote{Postdoc, L3S Research Center, Appelstr. 9a, 30167 Hannover, ildar.baimuratov@l3s.de}}
\affil{L3S Research Center, Hannover, Lower Saxony, 30167, Germany}
\affil{Cluster of Excellence SE²A – Sustainable and Energy-Efficient Aviation, Technische Universität
Braunschweig, Germany}

\author{Ludvig Knöös Franzén\footnote{Assistant Professor, Division of Fluid- and Mechatronic Systems, SE-581 83 Linköping, Sweden, ludvig.knoos.franzen@liu.se}}
\affil{Linköping University, Linköping, Östergötland, 58183, Sweden}

\author{Ingo Staack\footnote{Professor, Overall Aircraft Design, Hermann-Blenk-Str. 35, ingo.staack@tu-braunschweig.de} and Ulrich Römer\footnote{Professor, Modeling of Complex Systems under Uncertainty, Langer Kamp 19, u.roemer@tu-braunschweig.de}}
\affil{Technische Universität Braunschweig, Braunschweig, Lower Saxony, 38106, Germany}
\affil{Cluster of Excellence SE²A – Sustainable and Energy-Efficient Aviation, Technische Universität
Braunschweig, Germany}

\author{Sören Auer\footnote{Professor and Director, Data Science and Digital Libraries, Welfengarten 1B, 30167 Hannover, auer@tib.eu}}
\affil{TIB – Leibniz Information Centre for Science and Technology and L3S Research Center, Hannover, Lower Saxony, 30167, Germany}
\affil{Cluster of Excellence SE²A – Sustainable and Energy-Efficient Aviation, Technische Universität
Braunschweig, Germany}

\begin{document}

\maketitle

{
\small
\glsnogroupskiptrue
\printglossary[style=mcolindex,type=\acronymtype,nonumberlist]\itemsep0pt
}
\section{Introduction}
The aerospace industry operates at the frontier of technological innovation, combining complex systems and cutting-edge research to achieve advancements in design, manufacturing, and operational efficiency. With the current trend towards new system architectures, including novel
technologies, the complexity of aircraft design and its verification and validation is significantly growing and can no longer be handled with traditional engineering approaches. At the same time, the aerospace industry is known for high standards regarding safety and reliability, leading to rather incremental design evolution, where radical new configurations and products are rare. In this environment, with an enormous potential for re-use and adaptation of existing solutions and methods, it is not surprising that
\gls{kbe}~\cite{Stjepandić2015} has been applied for decades.  

\begin{figure}[htb!]
    \centering
    \includegraphics[width=0.95\linewidth]{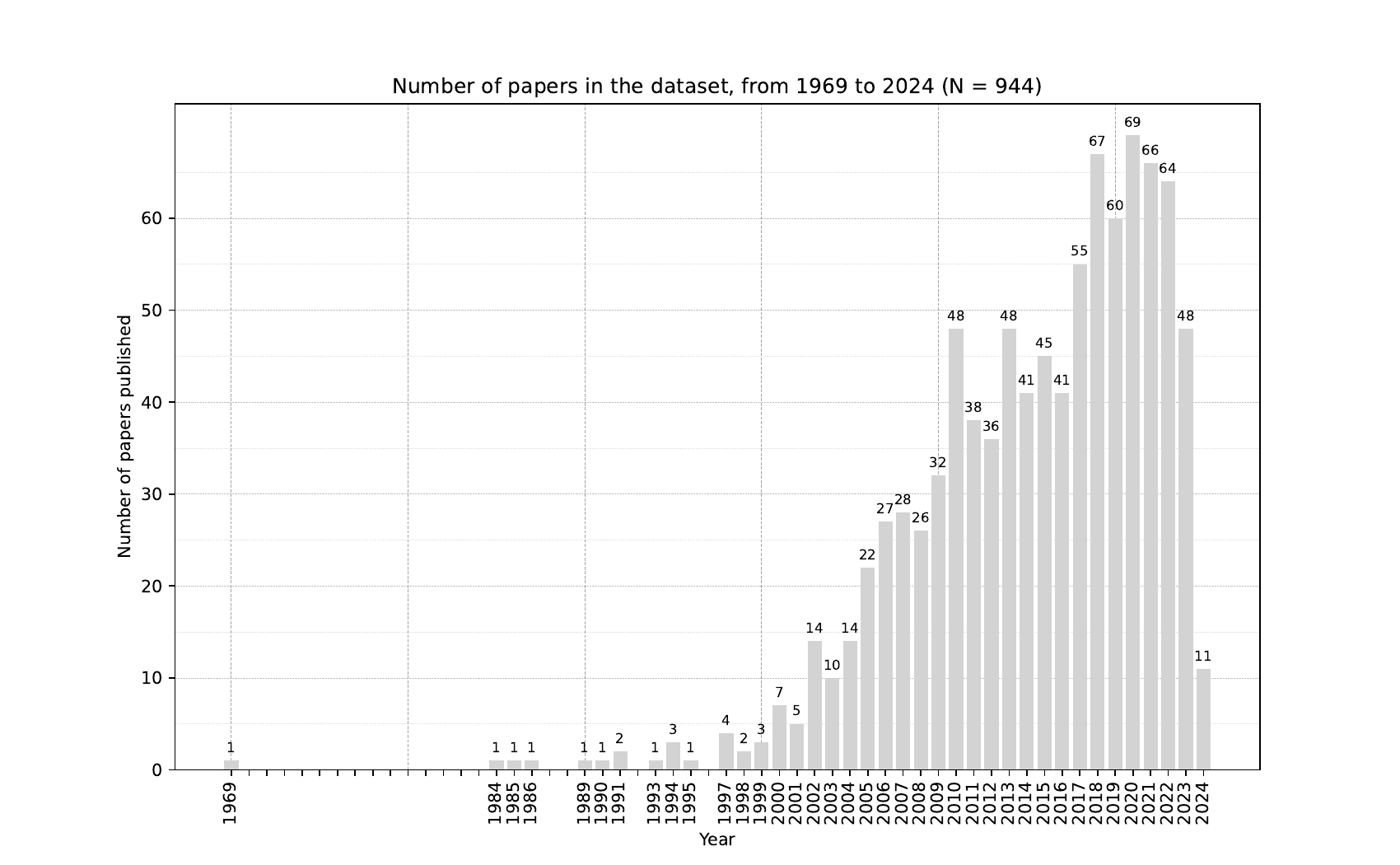}
    \caption{The 944 unique available papers to our knowledge-based aerospace engineering literature review, plotted over their publication year. A rise in relevant papers in recent years is noticeable.}
    \label{fig:year_processes}
\end{figure}

\gls{kbe} is an engineering methodology that focuses on capturing, structuring, and automating engineering knowledge to support decision-making and improve efficiency.
Nowadays widely used within \gls{cad} and design automation, traditional rule-based \gls{kbe} can be applied to any engineering domain in which data or information may be reused in a narrow or broader context. Examples range from structural design information reuse \cite{la_rocca_FE-models_2007}, requirement engineering \cite{boden2024automating}, and system component and architecture design \cite{van2008knowledge, staack2013systemKBEdesign}.
An overview of \gls{kbe} methods and tools is given in \cite{reddy_knowledge_2015, quintana2015transformingKBE}. 
\secref{fig:year_processes}
highlights the increasing spread throughout the aerospace domain in recent years with a trend to extend \gls{kbe} towards knowledge capture and management not solely focusing on explicitly stated information but making use of automatically extracted metadata from various sources, see e.g., \cite{schlichting2024knowledge}.
While conventional \gls{kbe} is well-suited for traditional evolutionary design engineering, the conventional \gls{kbe} methods have shortcomings with radically new design challenges, that are imposed by the need towards sustainable aviation \cite{reddy_knowledge_2015, matt_industrial_2023}.
Here, the potential of \gls{kbe} with regards to flexible data reuse and information needs to be unlocked.
These needs imply findability, accessibility, interoperability and reusability (\glsunset{fair}\gls{fair}) ~\cite{wilkinson2016fair} of aerospace engineering knowledge. 
This \gls{km} process, focusing on extraction, creation and utilization of machine-readable knowledge representation, is referred to as knowledge engineering.

The objective of this study is to identify and examine state-of-the-art knowledge management practices in the field of aerospace engineering. As outlined in \secref{sec:method}, we performed a preliminary literature review in advance to refine the objective and methodology,
leading to the formulation of three research questions:
\begin{itemize}
	\item \textbf{RQ1:} By which means, such as knowledge representations and schemas, do aerospace engineers externalize, such as formalize and visualize, knowledge? 
	\item \textbf{RQ2:} By which means, such as knowledge bases and services, do aerospace engineers utilize, such as organize and interface with, explicit knowledge?
	\item \textbf{RQ3:} By which means, such as tools and methods, do aerospace engineers exchange, such as transfer and distribute explicit knowledge?
\end{itemize}
Answering these questions, we aim to provide a recommendation framework for the further adoption and improvement of knowledge management practices. 

In order to achieve this objective, the systematic literature review methodology \gls{swarm-slr}~\cite{wittenborg_swarm-slr_2024} was used. 
This approach offers guidance and state-of-the-art tool support for conducting systematic literature reviews, while simultaneously producing \gls{fair} artifacts.
With the help of this streamlined workflow, we proceeded to survey over 1,000 articles aiming to answer our research questions.
A thorough manual review of 245 computationally recommended articles resulted in 164 articles included for dedicated analysis.
With this approach, we collected over 700 knowledge-based aerospace-engineering-related items and classified them into several categories, such as processes, tools, and data, which are discussed in the sections \nameref{sec:processes}, \nameref{sec:tools} and \nameref{sec:data} respectively.
Our findings are discussed in \secref{sec:discussion}, and the work is concluded in \secref{sec:conclusion} with a final summary.

Our contributions include:
\begin{enumerate}
    \item A \gls{swarm-slr} of over 1,000 articles with qualitative analysis of 164 selected articles, supported by two aerospace engineering domain expert surveys.
    \item A knowledge graph of over 700 knowledge-based aerospace engineering processes, software, and data, formalized in the interoperable \gls{owl} and mapped to Wikidata\footnote{\url{https://www.wikidata.org/}} entries where possible. The knowledge graph is represented on the \gls{orkg}\footnote{\url{https://incubating.orkg.org/observatory/Knowledge_Based_Aerospace_Engineering}}, and an aerospace Wikibase\footnote{\url{https://aerospace.wikibase.cloud}}, for reuse and continuation of structuring aerospace engineering knowledge exchange.
    \item Our resulting intermediate and final artifacts of the knowledge synthesis, available as a Zenodo dataset\footnote{\url{https://zenodo.org/records/14790367?preview=1&token=eyJhbGciOiJIUzUxMiJ9.eyJpZCI6ImQzZTczMWY1LWI2NmMtNDhiNS04Y2NjLTMwOGRlMzdmOGM3NyIsImRhdGEiOnt9LCJyYW5kb20iOiJkZTExMGE0OGRkODExYTc3ODQ3NDM4Y2JiMjFjYmZiYiJ9.gs-5KInBiNzr8pSu-YT0FzcGuMc76_F-PHW2jlFn-niYjbFBRxTZHBwUfkLEnAYyRW2e0SNy98cfA8yYRZk5cA} (Note to the reviewer: this is a preview accessible via token, for camera-ready we will provide a public url)}.
\end{enumerate}
The findings of this review not only provide a detailed understanding of current practices but also pave the way for developing interoperable knowledge-sharing infrastructures. 
These contributions are anticipated to foster collaborative innovation, enhance operational efficiency, and uphold the aerospace industry's commitment to excellence and sustainability.

The description of this survey and development process is structured as follows:
Initially, the \secref{sec:related_work} and \secref{sec:background} will provide the required context for this work.
Following this, \secref{sec:method} will cover all information related to the survey design.
Consecutively, \secref{sec:results} covers the general results, while three dedicated sections discuss the findings for each partition of the survey:
\secref{sec:processes} describes the processes relevant to either aerospace or knowledge engineers,
\secref{sec:tools} describes the tools used to solve these tasks, and
\secref{sec:data} concludes on which data items, formats, and models are used and produced by these activities.
How these findings could be further developed will be discussed in \secref{sec:discussion}.
Finally, the \secref{sec:conclusion} concludes this article.

\section{Related Work\label{sec:related_work}}
\begin{figure}[htb!]
    \centering
    \vspace{-1cm}
    \includegraphics[clip,trim={1cm 10.5cm 1cm 10.5cm},width=.9\linewidth]{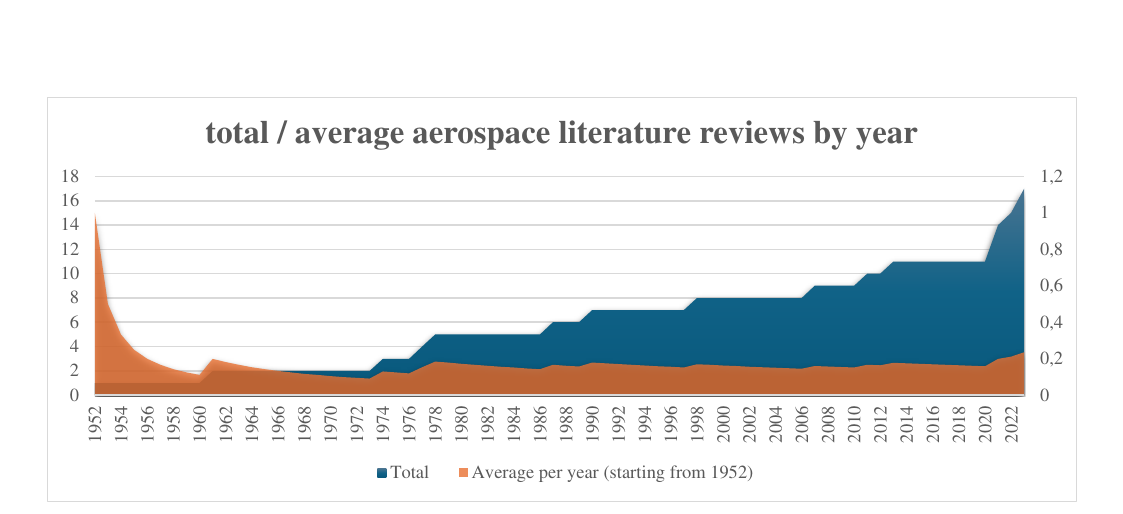}
    \caption{Development of \gls{aiaa} literature reviews over time. An increase since 2021 is noticeable.}
    \label{fig:Literature_reviews}
\end{figure}

Systematic literature reviews, which offer an unbiased and comprehensive overview of existing research, are rare in the field of knowledge-based aerospace engineering. Tallying the literature reviews and literature surveys present in the \gls{aiaa} search index, a new literature review is published only once every five years, starting from 1952, as illustrated in \secref{fig:Literature_reviews}. Using Google Scholar, OpenAlex, and \gls{aiaa}, we identified a total of 34 relevant literature reviews, which are curated in the Zotero library\footnote{\url{https://www.zotero.org/groups/5420201/sea-b-aerospace-knowledge-swarm-slr/collections/3J2G3DAG}}. However, most of these reviews focus on peripheral topics, lacking either an aerospace domain focus, a knowledge focus, or both, as depicted in \secref{tab:realted_work}.

\begin{table}[bth]
    \centering
    \caption{Related literature reviews by topics, in ascending similarity to our scope}
    \begin{tabular}{c l}
        \textbf{Source} & \textbf{Topic} \\
        \midrule
        \cite{torres_systematic_2021} & model management \\
        \cite{viana_metamodeling_2014} \cite{umland_multidisciplinary_2023} & \gls{mdo} \\
        \cite{kritzinger_digital_2018} \cite{enders_dimensions_2019} \cite{lim_state---art_2020} \cite{rathore_role_2021} \cite{kosacka-olejnik_how_2021} \cite{xames_systematic_2024} & digital twins \\
        \cite{hassan_application_2024} & application of \gls{ai} in aerospace engineering \\
        \cite{balaid_knowledge_2016} & knowledge maps \\
        \cite{schlenoff_literature_2013} & sensor ontologies in manufacturing \\
        \cite{cao_product_2012} \cite{kadiri_ontologies_2015} & \acrfull{plm}\glsunset{plm}, \cite{kadiri_ontologies_2015} particularly on ontologies in \gls{plm} \\
        \cite{rashid_toward_2015} & \acrfull{mbse}\glsunset{mbse} \\
        \cite{reddy_knowledge_2015} & \acrfull{kbe}\glsunset{kbe} \\
        \cite{kwakye_platform_2024} & platform health management in aerospace \\
        \cite{wang_knowledge_2018} & knowledge representations in concept development \\
        \cite{cerchione_using_2017} \cite{procko_leveraging_2022}  \cite{yao_influence_2023} & knowledge management, \cite{procko_leveraging_2022} specifically using linked data in the aerospace industry \\
        \cite{oprea_aspects_2021} & knowledge acquisition, exchange and collaboration in aerospace software development \\
    \end{tabular}
    \label{tab:realted_work}
\end{table}

Only three of these reviews focus on knowledge-based engineering. \citet{hassan_application_2024} examine the application of \gls{ai} in aerospace engineering.
\citet{oprea_aspects_2021} discusse knowledge acquisition, exchange standards, and multidisciplinary collaboration in software development. The most relevant work to our focus on knowledge-based engineering is \citet{procko_leveraging_2022}: \textit{``Leveraging Linked Data for Knowledge Management: A Proposal for the Aerospace Industry''}.
To our knowledge, none of these reviews apply \gls{fair} principles for their resulting artifacts, which are primarily single PDFs, nor do they incorporate a software-assistant methodology such as \gls{swarm-slr}.
In contrast, our systematic literature review provides digital artifacts, including knowledge graphs and software, that allow researchers to access, reuse, and extend our findings.

\paragraph{Preliminary Work\label{sec:preliminary}}
To complement the related work, it is worth mentioning that this \gls{slr} succeeds preliminary work.
Within the German Cluster of Excellence ``\acrfull{se2a}''\glsunset{se2a}
the project ``B4.2 - Consistent Multilevel Model Coupling and Knowledge Representation in Multidisciplinary Analysis and Design''\footnote{\url{https://www.tu-braunschweig.de/se2a/research/projects/area-b/b42-consistent-multilevel-model-coupling-and-knowledge-representation-in-multidisciplinary-analysis-and-design}} conducted a preliminary literature review.
The aim was to answer questions such as ``How do aerospace engineers formalize and exchange domain knowledge?'', with a strong focus on knowledge representations in multidisciplinary design with high-fidelity or multi-fidelity modeling.
This review was conducted throughout 2023 and informed the early stages of the B4.2 project while simultaneously laying the groundwork for the \gls{swarm-slr}~\cite{wittenborg_swarm-slr_2024}.
Equipped with this enhanced workflow, a better understanding of the domain specific literature, a set of 1,102 documents, and early results, the preliminary literature review was concluded in late 2023 to make way for a fresh, systematically structured start in early 2024.
This article represents this succeeding \glsentrylong{slr}.

\section{Background}\label{sec:background}
This work discusses aerospace \gls{kbe} with a focus on knowledge engineering data and knowledge representation.
To introduce these topics briefly, we provide context on \nameref{sec:kbe_aerospace} and \nameref{sec:knowledge_engineering} to finally provide a clearer grasp on \nameref{sec:semantic_web_in_aerospace}.

\subsection{Knowledge-Based Aerospace Engineering\label{sec:kbe_aerospace}}
Knowledge-based engineering leverages expert knowledge, rules, and computational tools to enhance engineering processes and improve their outcomes. In the aerospace industry, the development and management of complex systems require structured methodologies and robust data frameworks to ensure efficiency, traceability, and compliance.
Central to these methodologies is the concept of a process, which defines the sequence of activities required to design, develop, and maintain such systems.

\subsubsection{The Digital Thread}
\acrfull{plm}\glsunset{plm} is a structured approach for managing the entire lifecycle of a product from its initial concept through design, production, operation, and ultimately its retirement phase \cite{PLM2024}.
It emphasizes the integration of data, processes, and stakeholders across all lifecycle stages to improve efficiency and informed decision-making. A key enabler of this connectivity is the digital thread. 
The digital thread represents a continuous flow of data and information across a system's lifecycle \cite{DigitalThread2023}. It thereby serves as a digital representation of the system, thus enhancing aspects such as traceability, risk management, and decision-making support. 
These improvements are enabled through the comprehensive overview that the digital thread offers from the first design stage to the system's eventual retirement and all stages there in between.
Together, a digital thread and knowledge-based engineering enable data-driven decision-making capabilities that facilitate information exchange, efficiency, and potential for innovation.
This also complements the procedural model known as the ``engineering V-model'' by strengthening its emphasis on continuous verification and validation throughout a system's lifecycle and providing a digital representation of it.  

\subsubsection{The V-Model}
\begin{figure}[bt]
    \centering
    \includegraphics[width=0.80\linewidth]{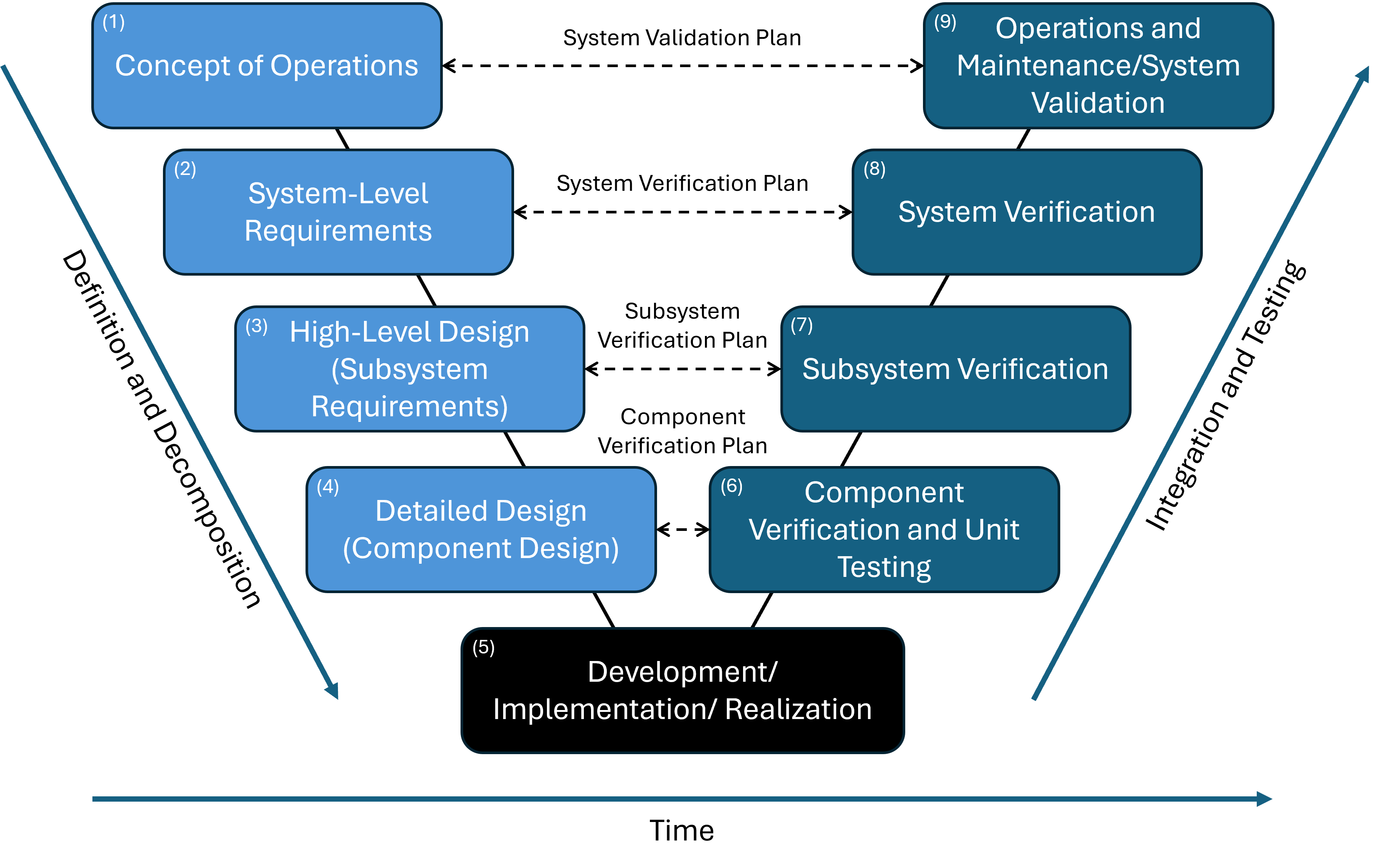}
    \caption{An illustration of the engineering V-model and its corresponding phases. Adapted from \cite{MITRE2014}.}
    \label{fig:VModel}
\end{figure}

The V-model, or engineering V-model, is a widely accepted systems engineering framework used to guide the development of complex systems, for example within aerospace applications~\cite{INCOSE2015, MITRE2014}.
The V-model emphasizes rigorous validation and verification processes throughout the product's lifecycle.
It is named for its characteristic ``V'' shape, where the left-hand side represents system definition and design phases, and the right-hand side represents integration and testing phases, ultimately ending in system deployment and maintenance. 
Figure \ref{fig:VModel} shows an illustration of the V-model structure with its corresponding phases. 

The emphasis on structured validation and verification processes from the V-model aligns well with the principles of knowledge-based engineering. However, the increasing complexity and scale of aerospace systems require data management-, integration-, and analysis tools that go beyond traditional frameworks, and this is where Semantic Web technologies come into the picture.

\subsection{Knowledge Engineering\label{sec:knowledge_engineering}}
With the exponential growth of data, the need for computational tools to manage, interpret, and utilize such information has become a topic of increasing importance. In engineering domains, especially aerospace, the complexity of systems and processes necessitates structured methodologies that enable efficient data handling and informed decision-making capabilities.
The \glsunset{fair}\gls{fair}~\cite{wilkinson2016fair} principles emphasize the capacity of computational systems to \textbf{F}ind, \textbf{A}ccess, \textbf{I}nteroperate, and \textbf{R}euse data with minimal human intervention.
Another initiative is Linked Data \footnote{\url{https://www.w3.org/DesignIssues/LinkedData.html}}, which aims to interlink data in the Web so it becomes possible to find them with semantic queries. To do that, \glspl{uri} are used to uniquely identify resources in the Web.
An example of the Linked Data implementation are DBpedia \cite{auer2007dbpedia} -- a project aiming to extract structured content from the information created in the Wikipedia project, and Wikidata \cite{10.1145/2629489} -- a collaboratively edited multilingual knowledge graph hosted by the Wikimedia Foundation. Finally, Google's \textit{Knowledge Graph} \footnote{\url{https://blog.google/products/search/introducing-knowledge-graph-things-not/}} became a common term for graph knowledge bases.

One of the technologies supporting \gls{fair} principles and Linked Data is the Semantic Web \cite{semantic_web}, which 
embeds machine-readable semantics into data through standards such as the \gls{rdf} and the \acrfull{owl}.

\begin{figure}[hbt]
    \centering
    {
    \includegraphics[width=0.7\textwidth]{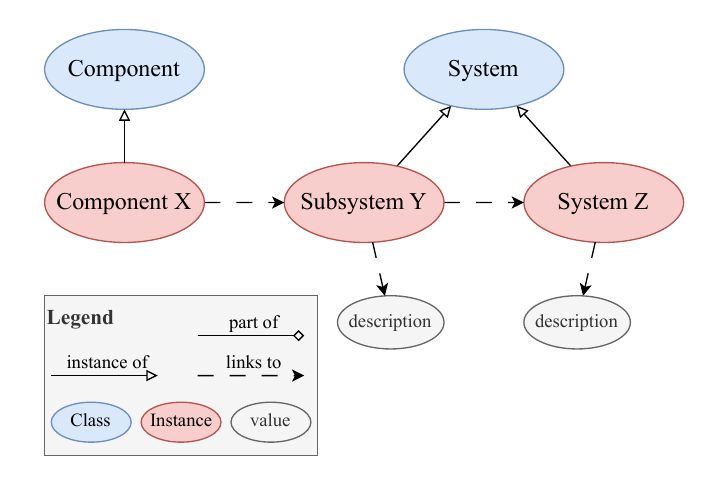}
    }
    \caption{Simple knowledge graph example, derived from \cite{meng_fault_2023}}
    \label{fig:Triple example}
\end{figure}

\paragraph{\gls{rdf}} 
is a framework for expressing information about resources. Resources can be anything, including documents, people, physical objects, and abstract concepts. \gls{rdf} is intended for situations in which information on the Web needs to be processed by applications, rather than being only displayed to people. It also provides a common framework for exchanging information between applications without loss of meaning.
\gls{rdf} allows us to make statements about resources. A statement always has the following structure: $<subject> <predicate> <object>$. The subject and the object represent the two resources being related and the predicate represents their relationship. As \gls{rdf} statements consist of three elements, they are called triples. The ability to have the same resource in the subject position of one triple and in the object position of another one makes it possible to find connections between triples and build connected graphs. These graphs can be queried with the \gls{sparql}.
Figure \ref{fig:Triple example} shows an example of triples that together form a small knowledge graph.

\begin{figure}[htb]
    \centering
    \includegraphics[width=0.6\linewidth]{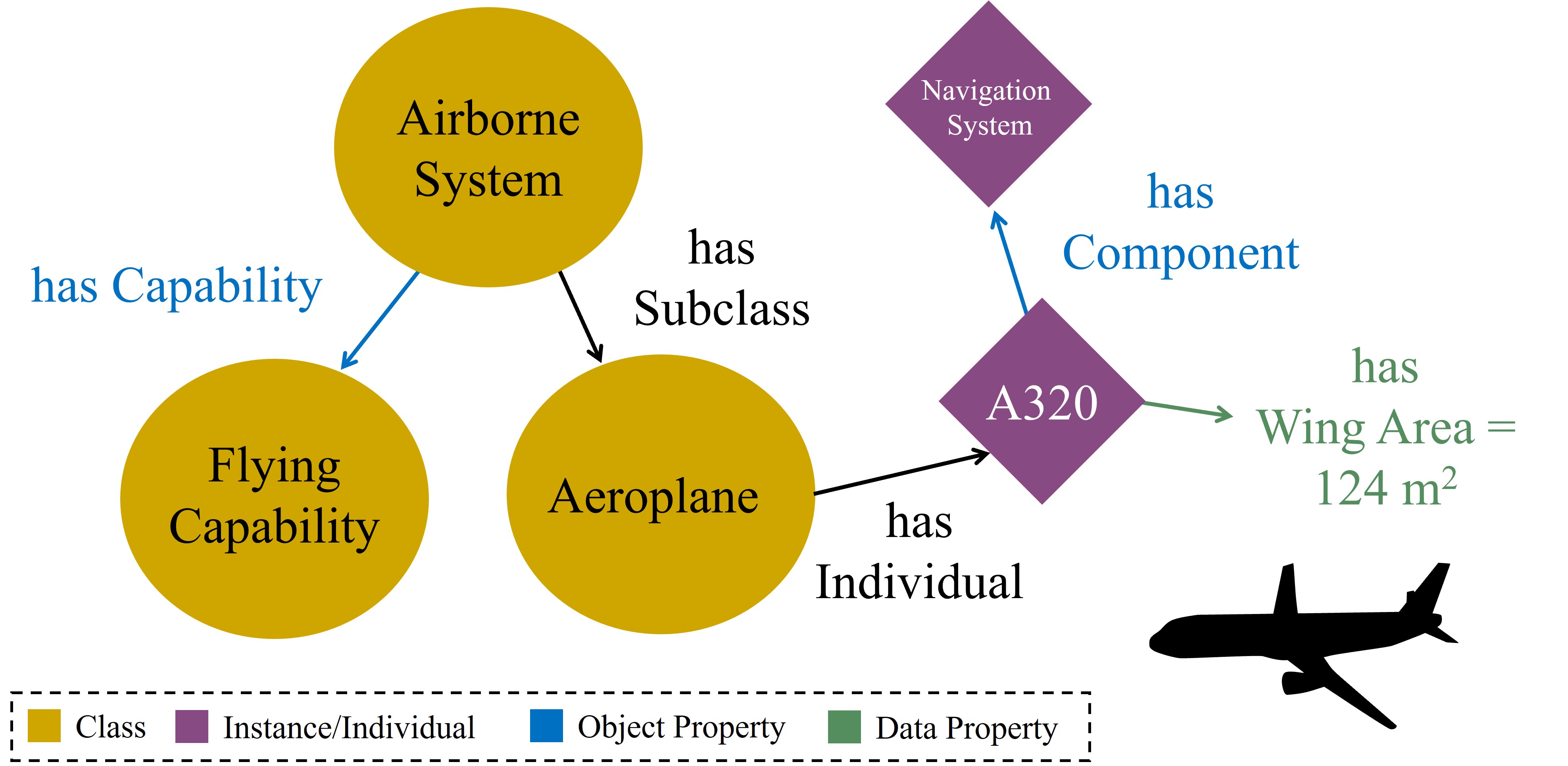}
    \caption{An aerospace example that illustrates some of the key aspects of an ontology.}
    \label{fig:Aircraft Ontology}
\end{figure}

\paragraph{\gls{owl}} is a vocabulary extension of \gls{rdf} that facilitates greater machine interpretability of web content. \gls{owl} can be used to explicitly describe semantics of data, resulting in an ontology.
Ontologies define \textit{Classes}, which unify certain \textit{Instances} under a common definition of shared attributes. Beside class membership relations, instances can be inter-connected via \textit{ObjectProperties}. Additionally, instances can have connections with non-object data such as integers, strings or similar via \textit{DataProperties}. To increase the human-readability, additional \textit{AnnotationProperties} can further specify information that is not structurally significant, such as a label or description.
Given such an ontology, reasoning engines can derive logical consequences from data, i.e. facts not literally present in the knowledge base, but entailed by the semantics.
Figure \ref{fig:Aircraft Ontology} shows an illustration of a small ontology within the aerospace domain.

Semantic Web technologies inherently promote findability, accessibility, interoperability, and reusability (\gls{fair}). They ensure that data is machine-readable, consistently structured, and easily discoverable by researchers and engineers.
These technologies facilitate the automation of repetitive design and analysis tasks through rule-based systems and AI-driven tools, streamlining workflows and fostering innovation.
In addition, as the Semantic Web is inherently distributed, ontologies must allow for information to be gathered from distributed sources. This is partly done by making the open-world assumption.
This assumption essentially means that just because something is not in the knowledge base, it does not mean that it is false; new information can always be added.
A number of different serialization formats exist for writing down \gls{rdf} graphs and ontologies: N-Triples, Turtle, \gls{json-ld}, \gls{rdfa}, RDF/XML, etc.
For a more detailed introduction of \gls{rdf} and \gls{owl} languages and their serialization we invite readers to read the RDF Primer\footnote{\url{https://www.w3.org/TR/rdf11-primer/}}and the OWL Language Overview from \gls{w3c}\footnote{\url{https://www.w3.org/TR/owl2-overview/}}.

\subsection{Semantic Web in Aerospace Engineering\label{sec:semantic_web_in_aerospace}}

Semantic Web technologies provide structured frameworks and tools that significantly enhance \gls{kbe} in the aerospace domain by improving the organization, accessibility, and automation of knowledge:
\begin{itemize}
    \item \textit{Data Interoperability:} \gls{rdf} and \gls{owl} enable the creation of ontologies that standardize how aerospace engineering knowledge is represented, thus ensuring that diverse systems can seamlessly exchange and interpret data.
    \item \textit{Knowledge Representation:} Ontologies define relationships between concepts in aerospace engineering, such as processes, tools, and data formats. This structured representation allows for detailed modeling of complex engineering workflows, improving the organization and retrieval of relevant knowledge.
    \item \textit{Enhanced Querying and Reasoning:} Using query languages like \gls{sparql}, Semantic Web technologies allow engineers to retrieve specific knowledge from large datasets efficiently. Reasoning engines can infer new knowledge from existing data, aiding decision-making and problem-solving.
    \item \textit{Linked Data:} By linking aerospace engineering data with external resources (e.g., Wikidata or DBpedia), Semantic Web approaches provide richer context and enable cross-domain knowledge integration. This interconnected data can support multidisciplinary collaborations crucial in aerospace projects.
\end{itemize}
This capability is particularly relevant in knowledge-based engineering, where domain-specific knowledge, automation, and optimization drive the design and lifecycle management of complex systems.
Both \gls{rdf} and \gls{owl} have been used in different works related to aerospace.
An example of this is \cite{KnöösFranzenThesis2023} where ontologies with \gls{dl} reasoning capabilities have been used to represent and process design spaces for an aerospace system of systems at an early design stage.
Other examples reach from \citet{dadzie_applying_2009} investigating the potential of using \gls{rdf} and \gls{owl} for knowledge sharing in aerospace engineering to \citet{rovetto2017ontology} to reliably using these for space data \gls{km} while extending towards \gls{kif} and \gls{clif}. Our systematic literature review primarily focuses on such studies that apply Semantic Web technologies in knowledge-based aerospace engineering.

\section{Review Methodology: A SWARM-SLR Approach\label{sec:method}}
Conducting a \acrfull{slr}\glsunset{slr} for a domain as complex and multidisciplinary as aerospace engineering requires a robust and efficient methodology. 
This review was conducted according to the \acrfull{swarm-slr}\glsunset{swarm-slr}~\cite{wittenborg_swarm-slr_2024}.
This methodology is particularly well-suited for knowledge-based aerospace engineering, as it addresses the challenges of managing diverse sources, ensuring comprehensive coverage, and maintaining consistency in data extraction and synthesis. 
The \gls{swarm-slr} approach is designed to streamline the review process through its four distinct stages: Planning, Searching and Selecting, Analyzing and Synthesizing, and Reporting, as depicted in \secref{tab:task_overview}.
These stages incorporate iterative refinement, enabling researchers to adapt the methodology to the evolving objectives and findings of the review.

\begin{table}[h]
    \centering
    \caption{Eight tasks conducted over the four stages of this \gls{swarm-slr}.}
    \begin{tabular}{c c}
        \multirow{2}{*}{Stage I} & \nameref{sec:task1}\\
        & \nameref{sec:task2}\\
        \midrule
        \multirow{3}{*}{Stage II} & \nameref{sec:task3}\\
        & \nameref{sec:task4}\\
        & \nameref{sec:task5}\\
        \midrule
        \multirow{2}{*}{Stage III} & \nameref{sec:task6}\\
        & \nameref{sec:task7}\\
        \midrule
        Stage IV & \nameref{sec:task8}\\
    \end{tabular}
    \label{tab:task_overview}
\end{table}

According to the \gls{swarm-slr}, each task has certain steps \textit{(and results)}. This section will go over each task and step to understand the approach, while the achieved results are generally discussed in the consecutive section. 

\subsection{Task 1: Planning a Review\label{sec:task1}}
Initiating a literature review generally begins with an imprecise inqury: 
\begin{enumerate}
    \item Formulate research interest {
        \it (explicit informal research interest)
    }
\end{enumerate}
As already described, the preliminary \gls{slr} significantly informed this first task and enabled a very clear outline in advance.
The abstract interests, such as ``Which services, software or algorithms that interface, utilize or visualize knowledge bases do aerospace engineers use/suggest?'', were already very close to the final research questions.

\subsection{Task 2: Defining a Scope\label{sec:task2}} 
As the final task of Stage I, this task concludes the definition of what this particular systematic literature aims to achieve:
\begin{enumerate}
    \item Check for related research questions {
        \it (narrowed informal research interest)
    }
    \item Refine scientific interest {
        \it (specific research question)
    }
    \item Formulate Search Query {
        \it (preliminary weighted keywords and queries)
    }
    \item Refine with related literature {
        \it (weighted keywords and refined queries)
    }
    \item Reevaluate with domain experts {
        \it (validated weighted keywords and queries)
    }
\end{enumerate}

Related research questions were continuously checked for during the preliminary work.
With that information, the scientific interest was eventually refined to three final research questions:

\begin{itemize}
    \item 
    \textbf{RQ1:}
    By which means, such as \textit{knowledge representations} and \textit{schemas},
    do aerospace engineers
    \textit{externalize}, such as \textit{formalize} and \textit{visualize},
    knowledge?
    \item 
    \textbf{RQ2:}
    By which means, such as \textit{knowledge bases} and \textit{services},
    do aerospace engineers
    \textit{utilize}, such as \textit{organize} and \textit{interface with},
    explicit knowledge?
    \item 
    \textbf{RQ3:}
    By which means, such as \textit{tools} and \textit{methods},
    do aerospace engineers
    \textit{exchange},
    such as \textit{transfer} and \textit{distribute}, 
    explicit knowledge?
\end{itemize}
These three research questions provide structured access to the knowledge-based engineering means and activities, selected as representatives of a larger relevant set that remains relevant as additional keywords.
Each of these questions was accompanied by an extensive list of weighted keywords and a search query.
For example, research question one
was associated with the keywords depicted in \secref{col:questions1} and the search query shown in \secref{col:query1}.
\begin{multicols}{3}
    [\captionof{lstlisting}{\textbf{SWARM-SLR readable representation of RQ1:} By which means, such as knowledge representations and schemas,
    do aerospace engineers
    externalize, such as formalize and visualize,
    knowledge?}]
    \begin{lstlisting}[
    label=col:questions1,
    basicstyle=\footnotesize
    ]
    ### aerospace engineering
    aerospace engineer:: 10
    aerospace engineering:: 10
    airspace engineer:: 10
    airspace engineering:: 10
    airplane engineer:: 10
    airplane engineering:: 10
    aironautic engineer:: 10
    aironautic engineering:: 10
    aircraft engineer:: 10
    aircraft engineering:: 10
    spacecraft engineer:: 10
    spacecraft engineering:: 10
    avionics engineer:: 10
    avionics engineering:: 10
    aviation engineer:: 10
    aviation engineering:: 10
    
    ### knowledge representation
    knowledge representation:: 10
    Knowledge graph:: 10
    Ontology:: 10
    information representation:: 8
    Taxonomy:: 8
    Semantic Web:: 8
    knowledge engineering:: 7
    schema:: 7
    expert system:: 6
    knowledge:: 5
    data representation:: 4
    semantic:: 4
    information:: 3
    repository:: 3
    documentation:: 3
    graphic:: 3
    index:: 3
    formal:: 2
    standard:: 2
    format:: 2
    structure:: 2
    graph:: 2
    plot:: 2
    mapping:: 2
    wiki:: 2
    technique:: 2
    classification:: 2
    categorization:: 2
    data:: 1
    map:: 1
    computer-aided design:: 1
    CAD:: 1
    computer-aided design:: 1
    virtual reality:: 1
    VR:: 1
    simulation:: 1
    platform:: 1
    report:: 1
        
    ### externalize
    externalize:: 10
    represent:: 10
    formalize:: 8
    visualize:: 8
    store:: 4
    structured:: 4
    articulate:: 3
    express:: 3
    manifest:: 3
    embody:: 2
    model:: 2
    document:: 1
    codify:: 1
    transfer:: 2
    explicit knowledge:: 8
    
    ### Formats
    RDF:: 4
    OWL:: 4
    XML:: 4
    CPACS:: 4
    SysML:: 4
    UML:: 4
    JSON-LD:: 3
    JSON:: 2
    YAML:: 2
    GraphML:: 1
    XSD:: 4
    \end{lstlisting}
\end{multicols}

\begin{lstlisting}[
    label=col:query1,
    basicstyle=\footnotesize,
    caption={\textbf{RQ1} Query with 929 Google Scholar results}
    ]
    "aerospace engineering" OR "aviation engineering" "Knowledge Representation" OR schema OR
    "Knowledge graph" externalize OR represent OR formalize OR visualize RDF OR OWL OR XML
\end{lstlisting}
Each research question was associated with one such search query, based on its weighted keyword list.
The keyword lists, without the \texttt{\#\#\# aerospace engineering} block which are equal among all research questions, are displayed in the appendix in \secref{col:questions2} and \secref{col:questions3}, alongside their respective queries in \secref{col:query2} and \secref{col:query3}.

Concluding this step, the weighted keywords and queries were evaluated by domain experts and deemed effective.
The search queries remained unchanged, but the keyword list and weights were continuously refined throughout the literature review to improve the recommendation further.

\subsection{Task 3: Search\label{sec:task3}}
Searching is a two-part, straight-forward task:
\begin{enumerate}
    \item Find resources {
        \it (preliminary document set)
    }
    \item Remove duplicates and find missing documents {
        \it (curated document set)
    }
\end{enumerate}
Following the \gls{swarm-slr} workflow, we used the three designed search queries on Google Scholar with the Zotero Plugin, resulting in 916, 509 and 30 results, providing gradually less results as Google Scholar limited the access over Zotero.
Exporting the list from Google Scholar directly somewhat compensated this limitation, but the provided overviews were not qualitatively or even quantitatively identical to the list available through the frontend, resulting in 892, 732 and 210 entries, respectively.
Combined with the 1102 entries from the preliminary work, this formed our document set of 4073.
After removing duplicates and finding available PDFs, 1029 documents remained for further processing. 

\pagebreak
\subsection{Task 4: Select\label{sec:task4}}
The selection task aims to identify relevant literature among the search results, requiring several steps to get there:
\begin{enumerate}
    \item Extract structured data from documents {
        \it (metadata;
        \it content;
        \it bag of words;
        \it contribution statements)}
    \item Calculate relational measurements within the document set {
        \it (document representation;
        \it similarity of documents within the document set)}
    \item Identify documents relevant for the research questions {
        \it (relevance of documents for research question)}
\end{enumerate}
The \gls{swarm-slr} algorithm processed the documents and scored them according to the weighted keywords from \nameref{sec:task2}.
While each RQ had its keyword set weighting in the aerospace aspect, it was made apparent that documents that sufficiently scored in other keywords could be classified as relevant without being related to the aerospace domain.
To address this, the aerospace aspect itself became an artificial fourth ``research question'' consisting of only the \texttt{\#\#\# aerospace engineering} block as shown in \secref{col:questions1}.
This score was used as a visual and algorithmic indicator to discourage investing time in non-aerospace literature.
Since this task is where Systematic Review Automation (SRA) is currently at its strongest, little further intervention was required to progress to the next task. 

\subsection{Task 5: Evaluate\label{sec:task5}}
Evaluation is the intermediate task between select and analysis, ensuring no categorically excludable artifact progresses to analysis.
\begin{enumerate}
    \item Remove out-of-scope document subset {
        \it (in-scope document set)
    }
    \item Evaluate relevancy measurement {
        \it (selection approval)
    }
    \item Classify document subsets {
        \it (literature subsets)
    }
\end{enumerate}

\begin{figure}[bt]
    \centering
    \includegraphics[width=1.\linewidth]{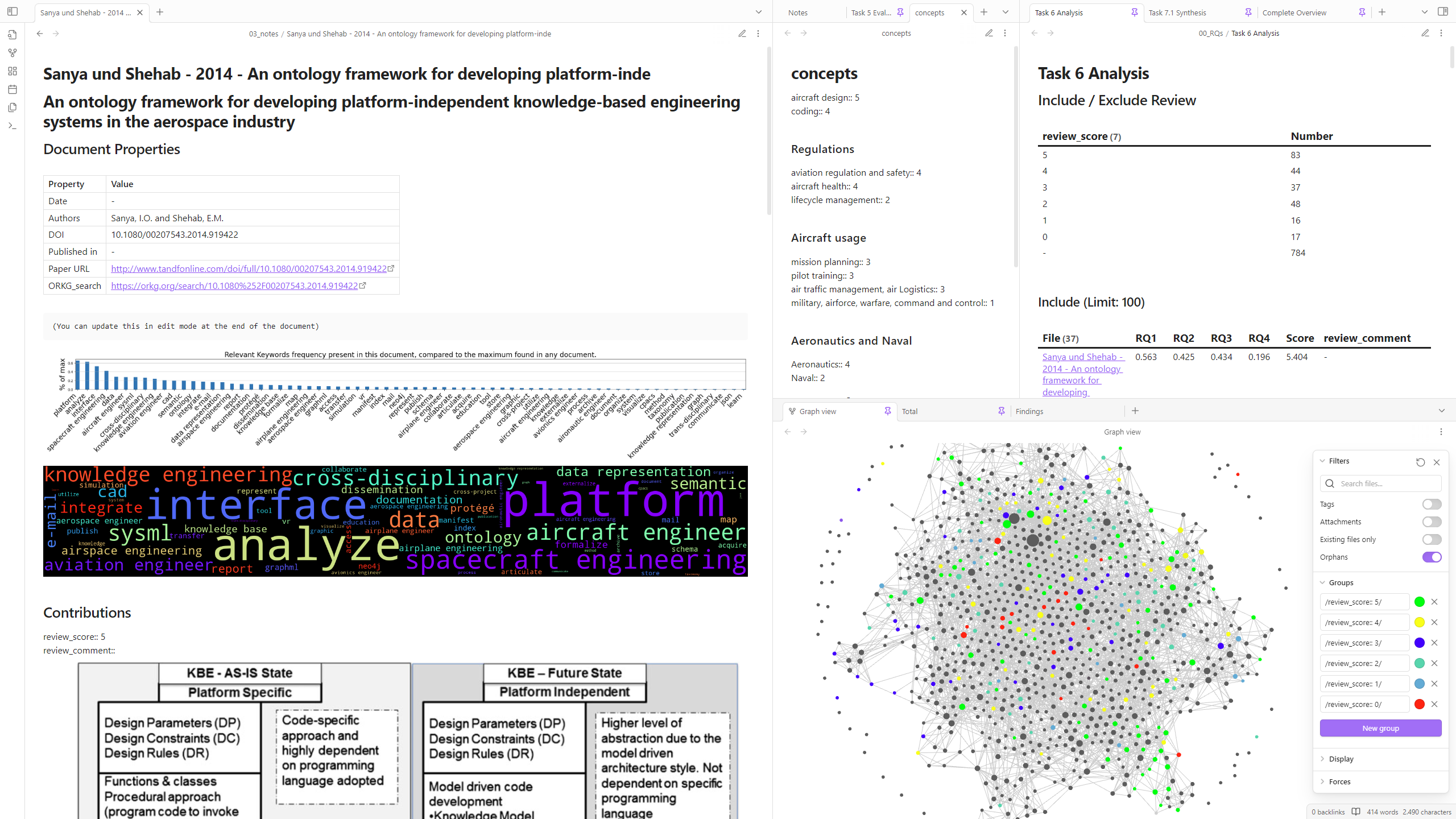}
    \caption{Obsidian\protect\footnotemark{} interface for \textit{Task 5: Evaluate}}
    \label{fig:task_5_obsidian}
\end{figure}

\secref{fig:task_5_obsidian} depicts the Obsidian based visual interface used in \textit{Task 5: Evaluate}.

A 5-score system was established, manually scoring each artifact already graded by the algorithm with a score from 5 (highly recommend) to 1 (reject), with an additional category 0 for categorically excluded artifacts such as duplicates, lists of abbreviations, university course structures, etc.
The evaluation surfaced the many different perspectives and domains that can mention aerospace terminology. We assigned different weights to each domain to facilitate the classification of in-scope and out-of-scope documents:

\begin{itemize}
    \item \textbf{5 strong match}: e.g. Aircraft design
    \item \textbf{4 weak match}: e.g. aerospace software development,
    aviation regulation and safety,
    aeronautics
    \item \textbf{3 borderline}: e.g. aircraft health,
    lifecycle management,
    mission planning,
    pilot training,
    naval,
    air traffic management, air logistics
    \item \textbf{2 weak mismatch}: e.g. e-commerce
    \item \textbf{1 strong mismatch}: e.g. military, air force, warfare, command and control
\end{itemize}

\pagebreak
\footnotetext{\url{https://obsidian.md/}}

\subsection{Task 6: Analysis\label{sec:task6}}
The analysis is a time-intensive manual process with a single step:
\begin{enumerate}
    \item Annotate resources {
        \it (contribution representation)
    }
\end{enumerate}
The qualitative analysis of the included and borderline papers was conducted by the first two authors of this work.
Starting from the highest rated documents, each reviewer annotated with an individual focus on certain research questions.
One reviewer focussed on how explicit knowledge is utilized and exchanged (RQ1 and RQ3), while the other reviewer paid particular attention on how explicit knowledge is formalized (RQ1).
Since both reviewers read the same papers, both could verify and complement each other's findings. This synergetic effect became particularly prevalent during \nameref{sec:task7}.

\subsection{Task 7: Synthesis\label{sec:task7}}
Given annotated resources, the knowledge can now be synthesized into a systematic overview:
\begin{enumerate}
    \item Compare contributions {
        \it (comparison)
    }
    \item Craft arguments based on findings {
        \it (document claims)
    }
    \item Critique the literature {
        \it (thesis claims)
    }
\end{enumerate}
After an initial pooling of the analysis results, we inquired with the help of \nameref{sec:interview} to facilitate a research-question- and domain-tailored \nameref{sec:kg_creation} process. 
We also utilized \nameref{sec:survey} to further strengthen this data acquisition and interconnection.
Finally, we conducted extensive
knowledge graph curation towards, among others, semantic disambiguation, decomposition, and ontology mapping.

\subsubsection{Expert Interviews\label{sec:interview}}
To streamline the annotation, a data model was designed to structure further synthesis. 
The first entries were sorted into a framework composed of categories such as processes, agents as well as data items, models, and formats.

Conclusively, a simple sentence covers all relevant classes:

Which 
Aerospace Engineering
\textbf{Process}
is completed by which
\textbf{Software}
using which
\textbf{Data}
in which
\textbf{Format}
and
\textbf{Schema}?

All instances and their relations relevant to answering our research questions can be sorted into one of these classes.
For example, RQ1 \textit{``By which means, such as knowledge representations and schemas,
do aerospace engineers
externalize, such as formalize and visualize,
knowledge?''}
asks which data items and data models are used for the respective processes. 

Below we provide examples for each class:
\begin{itemize}
    \item Process: problems, tasks or activities.
    \item Software: tools, applications or agents.
    \item Data item: image, document, or \gls{3d} model.
    \item Data schema: \gls{cpacs}, \gls{owl}, or Python.
    \item Data format: \gls{xml}, \gls{json}, or \gls{csv}.
\end{itemize}

These interviews grounded the theoretical synthesis in the domain.

\subsubsection{Expert Surveys}\label{sec:survey}
Just as \citeauthor{cook_quantifying_2013}~\cite{cook_quantifying_2013} complemented their well-known literature review with a climate expert survey, we reached out to aerospace engineers.
The initial idea was to mimic the \nameref{sec:interview}, present \secref{fig:architecture}, and (a subset of) the 500+ instances to complete and validate the findings.
Surveys were distributed to provide a secondary source of information for validating the findings.
With the help of the consulted engineers from said interviews, this scope was reduced, lowering the response threshold as far as possible:
\textit{I am working on \textbf{<Task>} using \textbf{<Software>} and \textbf{<Data Format>}}.
Data items and models could then be derived from the used software and formats.

The initial survey was extended and distributed again with an additional entry for the position of tasks, tools, and file formats along the engineering V-model phases to gain insight into where knowledge-based practices typically occur in aerospace applications.
Additionally, the survey served as an initial indication of how the classification of processes in the generated ontology could be supported by survey responses for future iterations of this study. 

\subsubsection{Knowledge Graph Creation\label{sec:kg_creation}}
To properly synthesize a knowledge graph, many additional processing steps are required:
First, many instances might be different representations of the same process, e.g. ``implementation'' and ``realization''
Others are very specific, e.g. ``collection and sharing of design information through a single source of truth (ssot)'', which is constructed from \textit{collection} of \textit{information} of \textit{design} by \textit{single source of truth}, including the same composition for sharing such information.
For comparison, the Wikidata entry ``\href{https://www.wikidata.org/wiki/Q15088675}{data curation (Q15088675)}'' is 
``\href{https://www.wikidata.org/wiki/Property:P279}{subclass of (P279)}''
``\href{https://www.wikidata.org/wiki/Q3007618}{content curation (Q3007618)}''
``\href{https://www.wikidata.org/wiki/Property:P642}{of (P642)}''
``\href{https://www.wikidata.org/wiki/Q42848}{data (Q42848)}''.
Curating the available data requires several knowledge engineering activities, such as merging duplicates, untangling composite instances, entity linking and human validation.

\paragraph{Statistical Analysis}
Co-occurrence and proximity of instances within our document set could be identified through statistical data analysis and used for identifying relationships between items of different classes, such as what software is used for what process, or what data formats are processed by what software. 
The software we developed extends \gls{swarm-slr} to calculate these statistics and is available on GitHub\footnote{\url{https://github.com/borgnetzwerk/tools/tree/main/scripts/SLR}}.

\subsection{Task 8: Reporting\label{sec:task8}}
The final task of the literature review is to report on it:
\begin{enumerate}
    \item Write the literature review {
        \it (review article)
    }
\end{enumerate}
Naturally, the article itself is the final, central artifact of a literature review.
But throughout a \gls{swarm-slr}, several intermediate artifacts are created.
Making these available to future research is also part of reporting.
To further describe these artifacts, their content and the knowledge within, this description of the methodology is followed by an overview of the general \nameref{sec:results}.

\section{Results\label{sec:results}}
\begin{figure}[bth]
    \centering
    \includegraphics[clip,trim={0.7cm 0.3cm 0.7cm 0.3cm},width=\textwidth]{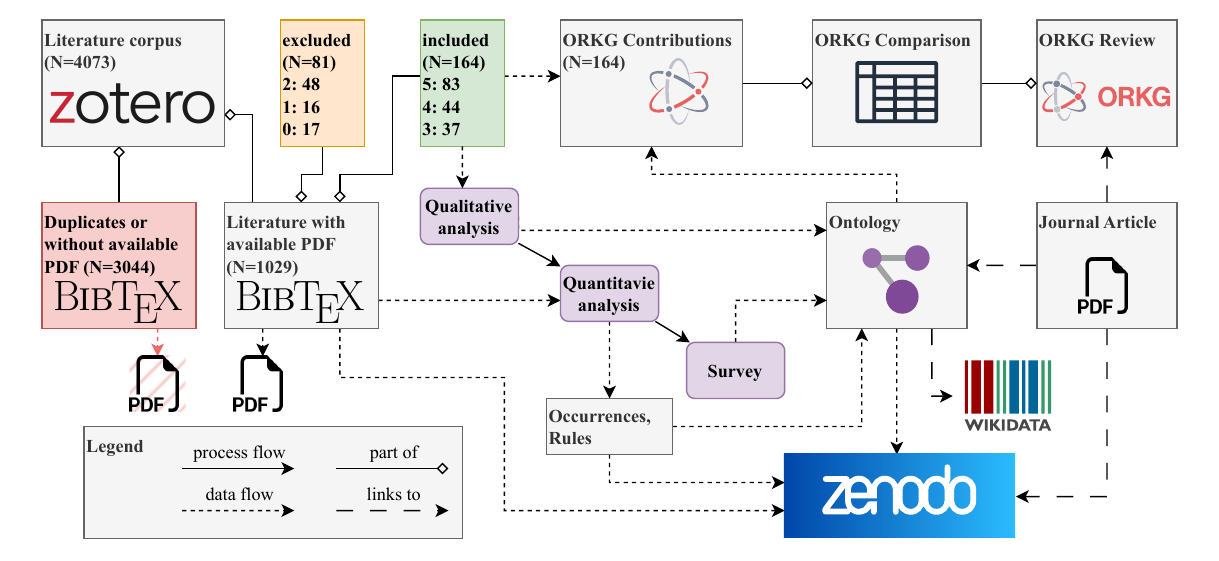}
    \caption{Artifact overview from a Zotero library to an ontology, linked to Wikidata and represented in the ORKG.}
    \label{fig:artifact_overview}
\end{figure}

This section illustrates the multitude of results of this SWARM-systematic literature review. 
\secref{fig:artifact_overview} presents an overview of how these particular results converge into this article while being independently accessible.
These artifacts include:
\begin{itemize}
    \item a Zotero library\footnote{\url{https://www.zotero.org/groups/5420201/se2a-b42-aerospace-knowledge-swarm-slr/library}} of all 4,073 retrieved and processed references, alongside the literature used for this article, i.e. related surveys,
    \item a Zenodo dataset containing
    \begin{itemize}
        \item our ontology with 782 individuals and relevant mappings to 154 contributions.
        \item a selection of intermediate data, such as occurrences and association rules.
    \end{itemize}
    \item several \gls{orkg} representations of the contributions, including
    \begin{itemize}
        \item 944 \gls{orkg} papers\footnote{\url{https://sandbox.orkg.org/observatory/KnowledgeBased_Aerospace_Engineering?contentType=Paper} (Note to the reviewer: for the reviewing phase, we keep them on the sandbox or incubating servers, for camera-ready they will be published on production)},
        \item an \gls{orkg} comparison\footnote{\url{https://incubating.orkg.org/comparison/R819284}} of the 164 included papers
        and their covered resources, namely processes, software as well as data items, schema and formats, 
        \item a review\footnote{\url{https://incubating.orkg.org/review/R819286}} compositing these \gls{orkg} artifacts alongside more details regarding search queries, keywords, etc. 
    \end{itemize}
\end{itemize}

Several subsections and dedicated finding sections will structure these results:
\begin{itemize}
    \item \fullref{sec:res_qual} gives insight into several most relevant of the 164 analyzed articles. 
    \item \fullref{sec:res_synt} presents the ontology derived from this analysis, including its further detailing through expert surveys and interviews.
    \item \fullref{sec:res_quan} discusses what is achieved with this ontology, ranging from a lightweight quantitative analysis over the entire retrieved literature to an \gls{orkg} comparison and review.   
    \item \fullref{sec:processes}, \fullref{sec:tools} and \fullref{sec:data} detail class-specific findings for the three major categories.
\end{itemize}

\subsection{Qualitative Analysis\label{sec:res_qual}}
After the documents were computationally scored in Task 4, the highest scored articles were manually evaluated as described in Task 5. 
In total, we classified the 245 documents as shown in \secref{tab:reviewed_documents}:

\begin{table}[bth]
    \centering
    \caption{Scores from the manual review of the 245 highest-rated documents}
    \begin{tabular}{c c c c c c}
        \textbf{5 strong include} & \textbf{4 weak include} & \textbf{3 borderline} & \textbf{2 weak exclude} & \textbf{1 strong exclude}& \textbf{0 categorical exclude}\\
        \midrule
        83
        & 44
        & 37
        & 48
        & 16
        & 17
    \end{tabular}
    \label{tab:reviewed_documents}
\end{table}

As described earlier, reviewers annotated the 164
included and borderline papers concerning their contributions to the three research questions.
A selection of particularly noteworthy literature includes the following:

    \citeauthor{sanya_challenges_2011} is likely one of the prime authors to highlight in this \gls{slr}, discussing semantic knowledge management challenges in aerospace~\cite{sanya_challenges_2011} and the potential of using ontologies there~\cite{sanya_ontology_2014} a decade ago. A set of noteworthy sentences warrant citation:
    \begin{quote}
         ``\textit{It was established that improving the structure and representation of knowledge would benefit and increase the competency of design engineering activities. [...]
         The authors suggest that something more clear and concise is needed beforehand, namely that knowledge is structured and organised in an appropriate manner with appropriate vocabulary. The authors argue that this structure is needed in order to bring forth the next generation of knowledge base building, and ontological engineering is a prerequisite for this vision.}''\\
         --\citeauthor{sanya_challenges_2011}~\cite{sanya_challenges_2011}
    \end{quote}
    They specifically mention that ``engineers often spend large amounts of time searching for a solution to problems that may have been solved.'', partically because ``engineers are sometimes reluctant to share documents'' or ``do not want to be burdened with the task of knowledge classification or annotating the semantic contents of a document''. The authors recommend ``acquiring a dedicated knowledge/ontological engineer that will be responsible for such an activity''. For details, reading the entire paper and ideally some consecutive work like \cite{sanya_ontology_2014} is highly recommended.
    
    The dissertation by \citet{xie_application_2013} provides a good introductory work as to what \gls{kbe}, particularly knowledge itself, may mean in the context of aerospace. The work also contains a review of systems classified by knowledge capture automation, data mining techniques deployed, origin and applications. Exactly such overviews were already required back then and are now facilitated through interoperable technologies, such as the \gls{orkg}, as also employed in this work later. Using this technology, their identified knowledge is reused and enhanced for future dynamic extension. They describe a gap for exactly that as their final key finding, ``to apply semantic representation to capture aerospace context in an integrated and open manner.''
    
    \citet{gonzalez_complex_2016} provides an overview on complex multidisciplinary system composition for aerospace vehicle conceptual design, where particularly the list of 126 \textit{Aircraft Synthesis Systems} (based on \cite{chudoba2001stability,huang_prototype_2005,coleman_aircraft_2010}) is of interest to this work. It is apparent that semantic ambiguity even occurs within collections, within the same departments, listing the abbreviation to both mean ``Multi-Disciplinary Integrated Design Analysis \& Sizing'' for DaimlerChrysler Military, as well as ``Multi-Disciplinary Integration of Deutsche Airbus Specialists'' for DaimlerChrysler Aerospace Airbus. Examples like these highlight the complexity (tool-assisted) entity recognition has to face when digesting such knowledge sources, where arguably not even humans could differentiate one MIDAS mention from the other, given that both existed as separate instances in the first place. 
    
    \citet{haney_data_2016} speaks of ``aerospace data-engineering'', continuing the aformentiond issue of polysemes and synonyms. Likely, the tasks of an aerospace data engineer are virtually identical to a aerospace knowledge engineer. Formalizing these ``said to be the same as'' relationships is equally important for maintaining interoperability in systems that do not use perfactly aligned vocabularies, as is also the case in the scope covered here. \citeauthor{haney_data_2016} is also a strong source for making the flow of knowledge and the benefit of capture and reuse tangible, spanning several illustrations of how accessible, actively used digital libraries enrich the engineering process.
    
    \citet{aigner_graph-based_2018} discuss \gls{kadmos}, \gls{cmdows} and \gls{rce} in the context of graph-based formulation and visualization of \gls{mdo} systems. As the nested abbreviations already suggest, these highly intertwined systems are very complex and already benefitting from a \gls{kbe} approach. It is a very good example of knowledge-based, interoperability-focussed system design, widely used throughout the AGILE project it was part of. Many related works of the project, such as \cite{baalbergen_advancing_2022} of the AGILE 4.0 followup project, further build on these interoperable frameworks.
    
    \citet{van_gent_composing_2017} discusses similar themes, tools and models, with focus on \gls{cpacs} as a central data model. 
    It is positioned as a central data model to reduce the complexity of \(\mathcal{O}(N^N)\), every system \(X_i\) developing its own adapter for every other system \(X_{j \neq i}\) and back, to 2 adapters converting data from \(X \to \text{\gls{cpacs}}\) and \(\text{\gls{cpacs}} \to X\)), \(\mathcal{O}(2N)\). Recommendations of such universal aerospace interchange formats go back to 2004~\cite{agrawal_web-based_2004}, potentially beyond.
    
    \citet{procko_leveraging_2022}, as teased by the related work section, is actually a literature review of knowledge management and knowledge collaboration systems in the aerospace and defense industries. They build upon everything mentioned up until now and propose an \gls{owl} based knowledge management system for the aerospace domain. Their work follows a very similar research question, but using a less systematic literature review approach on a smaller subset. They do, however, go much more into detail regarding the specific needs of the aerospace domain. They, too, identify a quote-worthy sentiment:
    \begin{quote}
        ``\textit{In addition to these challenges, the aerospace and defense industries are perhaps the most governed of all sectors, being regulated by a plethora of agencies, both national and regional [22].
        \gls{km} is, put simply, about providing knowledge to people quickly; yet the tight regulations of these industries may hamper effective knowledge dissemination.}''\\
         --\citeauthor{procko_leveraging_2022}~\cite{procko_leveraging_2022} (and their source ``[22]'', \citet{harvey_knowledge_2005})
    \end{quote}
    They follow this up with by discussing knowledge protection:
    \begin{quote}
        ``\textit{As aerospace and defense sector project work is often performed in a collaborative effort between firms, there is concern that a firm’s specialized knowledge may be appropriated by the cooperating firm during collaboration, in a move of opportunistic advancing. “… firms experience a fundamental paradox: to gain the greatest benefits they must exchange information and knowledge with external parties yet, at the same time, they must protect themselves against knowledge appropriation. This dilemma is particularly acute in the aerospace sector where political imperatives strongly influence partner choice and collaborators are often strong rivals in other contexts” [68].}''\\
         --\citeauthor{procko_leveraging_2022}~\cite{procko_leveraging_2022} (and their source ``[68]'', \citet{jordan_protecting_2004})
    \end{quote}
    Conclusively, they advise for knowledge sharing despite the challenges, recommending accessible knowledge graphs for collaboration. 
    Since their recommendation is again exactly aligned with what our work has done, we can confirm that independently our conclusions overlap to recommend the aerospace domain to adopt \gls{fair} knowledge management, as exemplified by our work's immediate knowledge-graph-based results.

Every one of the 164 documents, whether mentioned here or not, contributed to the creation of the knowledge graph, whose synthesis is discussed in the following subsection.

\subsection{Synthesis\label{sec:res_synt}}
While we could already identify 533 total aerospace 
\gls{kbe} items via literature review, we validated our scope through expert interviews and surveys, as described in the previous chapter. 
The results are displayed in the following subsections:

\subsubsection{Wikidata Linking\label{syn}}
Eventually, instances were manually mapped to Wikidata items for disambiguation, following the Linked Data approach.
For example, \textit{Cypher} might mean one of 607 different things, but in our knowledge graph is clearly defined by being mapped to \url{https://www.wikidata.org/wiki/Q16834355}. As another example, Wikidata contains 69745 file formats, many of which are likely versions of the same file format families.
The lexeme \textit{trace} alone has 41930 entries in which we need to identify \textit{TRACE}, a software used for \textit{computational fluid dynamics (CFD)}.

Our literature review revealed that 
40 \% of the 215 processes, 
37 \% of the 135 software, 
52 \% of the 65 data items,  
54 \% of the 68 data models,  
90 \% of the 50 formats
could be mapped to Wikidata entries.

The more universal the relevance of an instance, like \textit{XML} or \textit{optimization}, the easier their identification.
Remarkably, even largely discussed domain-specific instances like \textit{\acrfull{cpacs}} or \textit{\acrfull{mdo}} did not have Wikidata entries.
Others, like the \textit{National Airspace System (NAS)}, were identifiable via Wikidata, but could not be categorized into our scope of software, processes, and data, since they are instances of concepts outside these categories, and thus excluded from our ontology.

\subsubsection{Expert Interviewes}
\begin{figure}[bth]
    \centering
    \includegraphics[width=.9\textwidth]{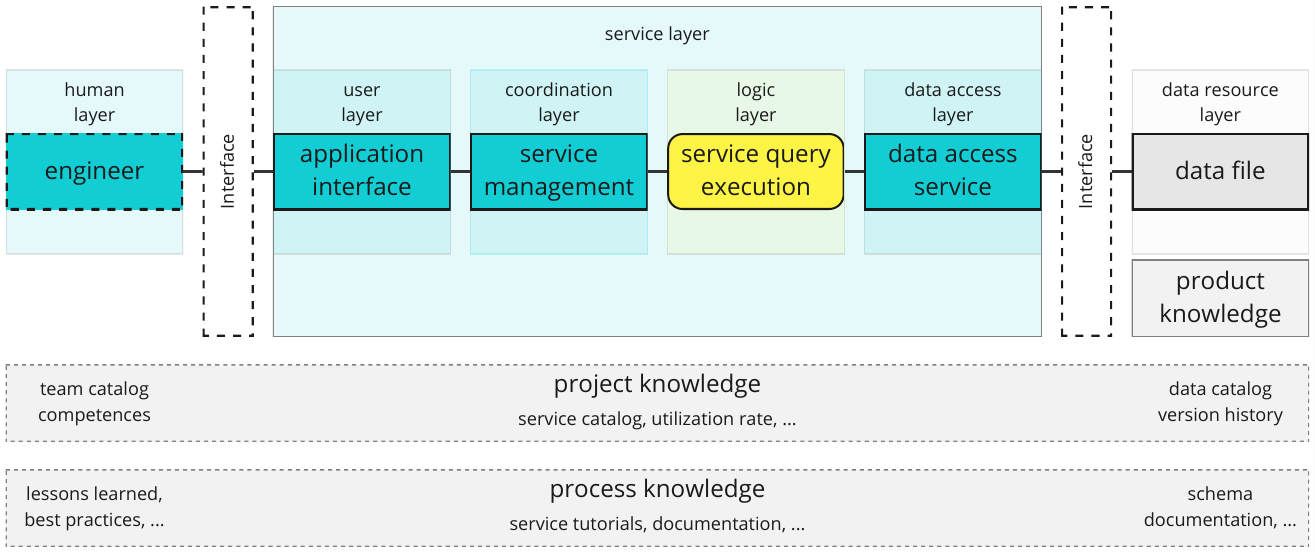}
    \caption{Knowledge architecture concept, based on surveyed literature (especially \citet{herrero_system_2022}). An engineer interfaces with a service, which in turn interfaces with explicit product knowledge in the form of data. Beyond product knowledge, knowledge-based engineering also covers project knowledge (to manage the product creation process) and process knowledge (to solve tasks efficiently).}
    \label{fig:architecture}
\end{figure}

We invited three aerospace engineers over three months for individual, intermediate evaluations. These interviews resulted in a refined concept depicted in \secref{fig:architecture}.
For example, \textit{agent} was reduced to a subcategory of agents, namely \textit{software}, which is technically more restrictive and potentially not a perfect fit, for example when considering non-software tools, services, or the involvement of a human actor in a process.
These compromises were taken to reduce an initial precise, but abstract ontology to a simple, yet sufficient model.

\subsubsection{Expert Surveys\label{sec:results_survey}}
A dedicated workshop for this systematic literature review was held within the \gls{se2a} cluster in the summer of 2024.
16 participants were introduced to the survey, approach, and results, and were offered to contribute to the survey.
14 responses were collected that day.
The survey was then distributed to the entire cluster of over 200 engineers working in the aerospace domain.
Over the following three months, three additional responses were collected.
The second survey included a brief description of the V-model phases for the participants' reference and was distributed to additional universities, such as Delft University of Technology (\gls{tu} Delft) and Linköping University (LiU), industry partners, and again in the \gls{se2a} cluster.
Ultimately, this survey received answers from a total of 14 engineers. 
These answers contribute an extension to the answers for the first survey while providing some additional insight into where knowledge-based practices typically are used along the engineering V-model phases.
The results from this analysis, along with a brief description of the engineering V-model, can be found in \secref{sec:processes}.

\subsubsection{Knowledge Graph Curation}
\begin{figure}
    \centering
    \includegraphics[clip, width=0.95\linewidth, trim={2cm 2cm 2cm 3.2cm}]{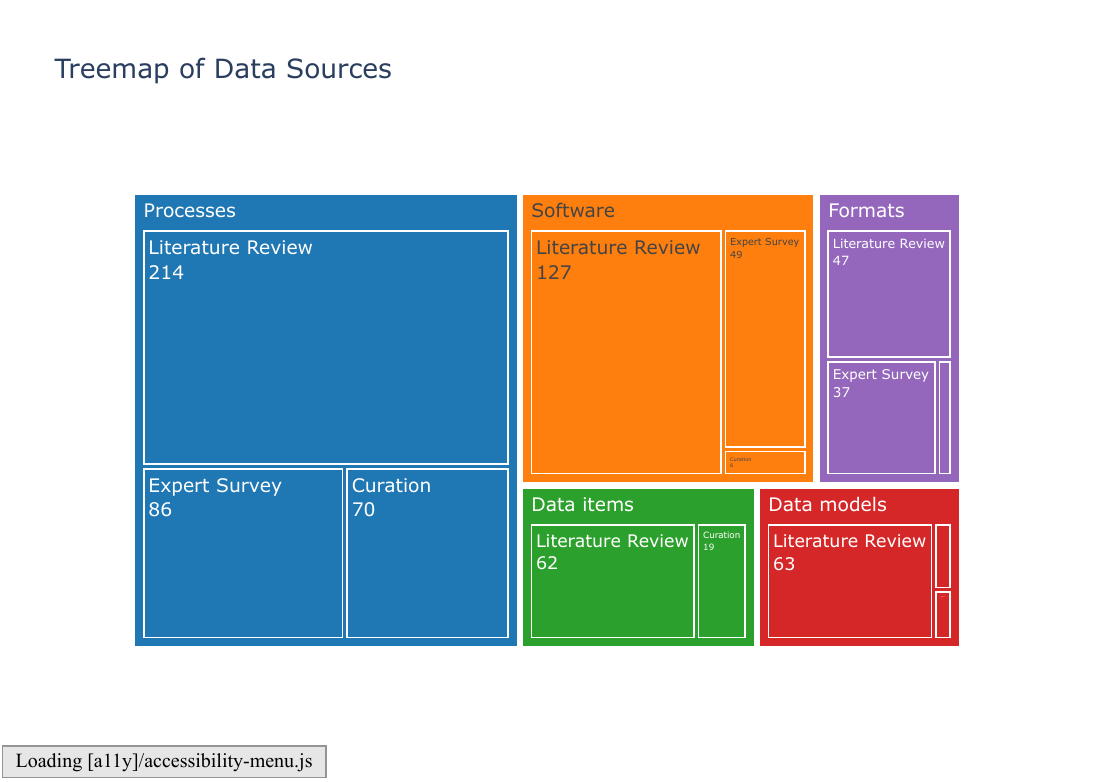}
    \caption{Treemap of the Source distribution of the knowledge graph items}
    \label{fig:source_treemap}
\end{figure}

As mentioned, 513 aerospace \gls{kbe} items were identified throughout the literature review.
With the expert surveys, this total grew to 695 instances.
The final, extensive curation raised the total to 792
instances as illustrated in \secref{tab:my_label} and \secref{fig:source_treemap}.
This table also specifies the percentages of items gained from each source. The literature review contributed the most items (65\%), proving most effective for data models (90\%) and least effective for formats (53\%). The expert survey added 23\% of the items, with the highest increase in formats (42\%) and no additional data items. Finally, curation accounted for 12\% of new items, with the largest gain in data items (23\%) and no significant increase in software, data models or formats.

\begin{table}[bth]
    \centering
    \caption{Sources of the aerospace engineering knowledge graph items}
    \label{tab:my_label}
    \begin{tabular}{l | c c c c c c}
         \textbf{Source} & \textbf{Total} & \textbf{Processes} & \textbf{Software} & \textbf{Data items} & \textbf{Data models} & \textbf{Formats} \\
         \midrule
         Literature review & 513 (65\%) & 214 (58\%) & 127 (70\%) & 62 (77\%) & 63 (90\%) & 47 (53\%) \\
         
         Expert survey & 182 (23\%) & 86 (23\%) & 49 (27\%) & 0 (0\%) & 4 (6\%) & 37 (42\%) \\
         
         Curation & 97 (12\%) & 70 (19\%) & 6 (3\%) & 19 (23\%) & 3 (4\%) & 5 (6\%) \\

         Total & 792 & 370 & 182 & 81 & 70 & 89 \\
    \end{tabular}
\end{table}

Though the performed curation is extensive, it still leaves room for further improvements, including inference, connection to existing ontologies, further interrelationship mapping, etc.
This curation is final for the scope of the present \gls{slr}, but just a first step for a future interoperable knowledge base for the aerospace engineering community.

\subsection{Statistical Analysis\label{sec:res_quan}}
Given a large, natural-language-processed document set of 1029 documents and a knowlege graph, we applied quantitative analysis techniques to derive insights.
These methods produced several visual artifacts based on statistical findings to further support curation.
This subsection provides an overview of the quantitative analysis techniues that have been used.
For access to the very large visualization, please see the Zenodo dataset\footnote{\url{https://zenodo.org/records/14790367?preview=1&token=eyJhbGciOiJIUzUxMiJ9.eyJpZCI6ImQzZTczMWY1LWI2NmMtNDhiNS04Y2NjLTMwOGRlMzdmOGM3NyIsImRhdGEiOnt9LCJyYW5kb20iOiJkZTExMGE0OGRkODExYTc3ODQ3NDM4Y2JiMjFjYmZiYiJ9.gs-5KInBiNzr8pSu-YT0FzcGuMc76_F-PHW2jlFn-niYjbFBRxTZHBwUfkLEnAYyRW2e0SNy98cfA8yYRZk5cA}}.

\subsubsection{Paper-Instance-Occurrence}
Using the extracted text and the exported ontology, literal matches in the text were identified.
Immediately, this enabled a binary occurrence matrix as a quantitative complement to the conducted qualitative analysis.
While the occurrence matrix, having dimensions of $793\times 1029$ (or $793\times 164$ if just for the reviewed papers) is too large to be displayed here, it is available in the appendix as \secref{fig:instance_instance_relative_co_occurrence_matrix}, and in full as part of the Zenodo dataset and on the \gls{orkg}, as further described in \fullref{sec:results_orkg}.

\begin{figure}
    \centering
    \includegraphics[trim={0 0 1 1cm},clip,width=.9\textwidth]{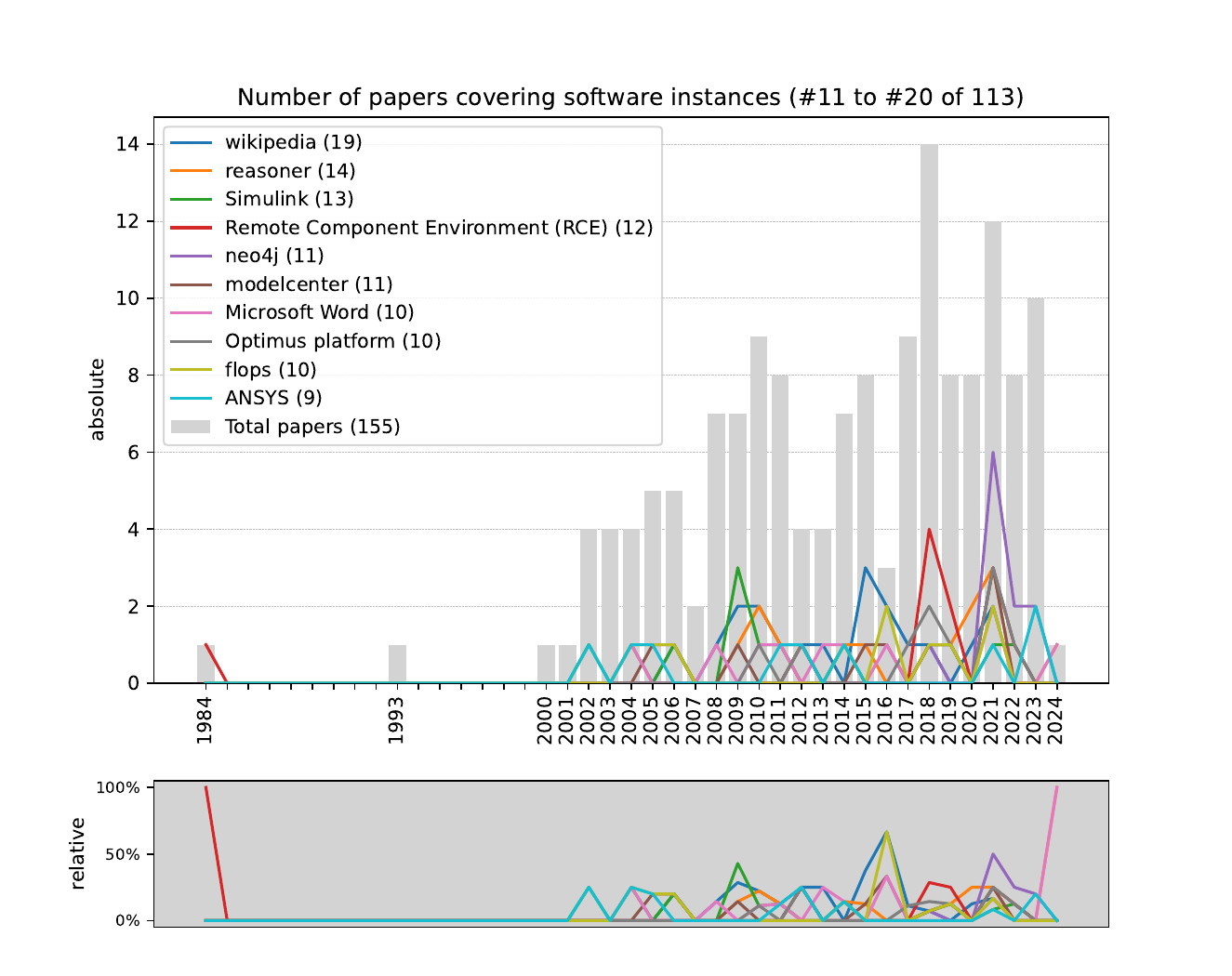}
    \includegraphics[trim={0 0 1 1cm},clip,width=.9\textwidth]{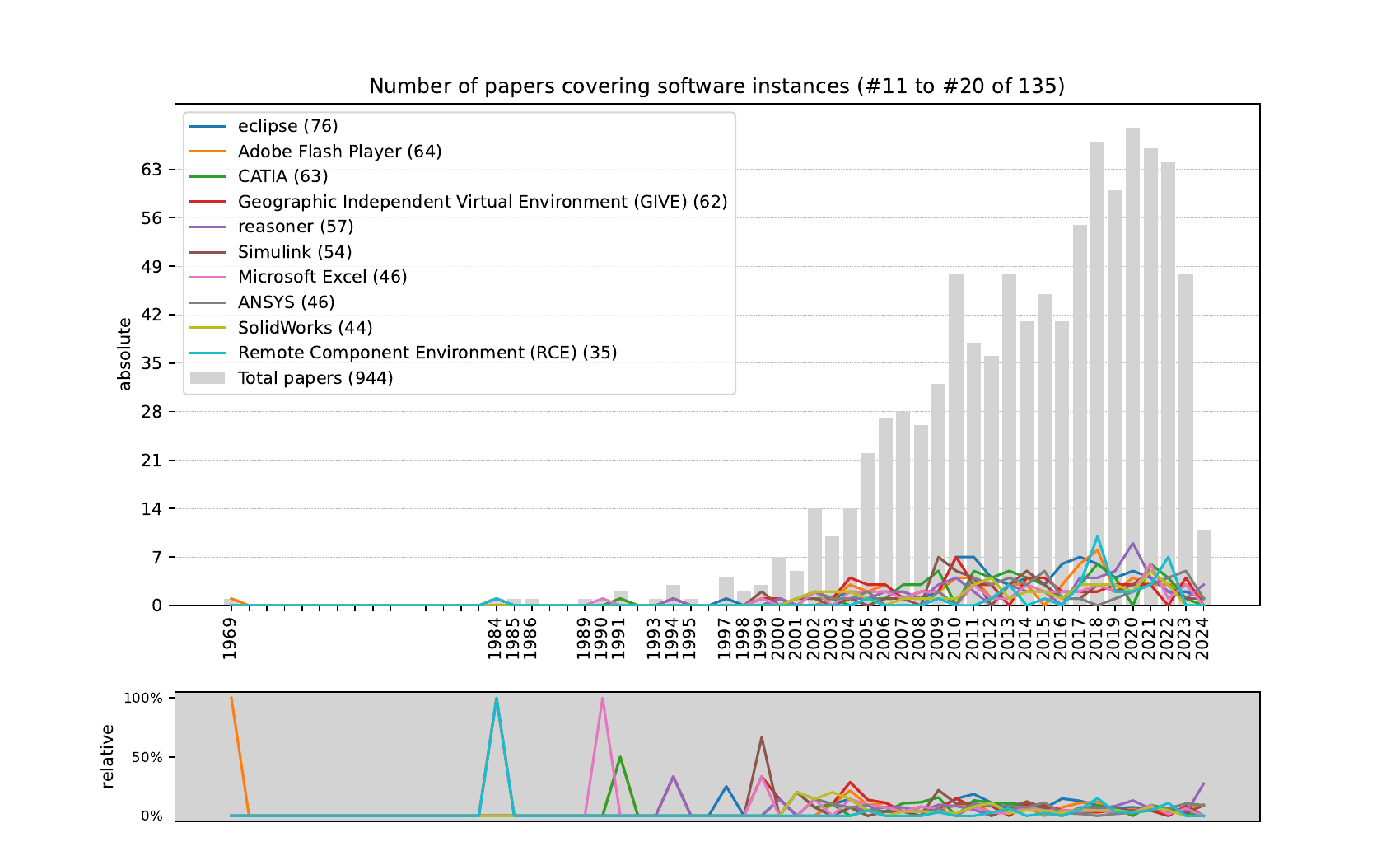}
    \caption{Instance occurrences in reviewed literature (above) compared to occurrences in entire corpus (below).}
    \label{fig:occurence_over_years}
\end{figure}

Having this occurrence data alongside the rich semantic data also allowed for dedicated, class- and year-specific overviews, as depicted in \secref{fig:occurence_over_years}.
This matching is by its very nature inaccurate.
False positives occur, not only when one MIDAS tool is misinterpreted for the other~\cite{gonzalez_complex_2016},
but also if an occurrence of the word \textit{``give''} is misinterpreted for the Geographic Independent Virtual Environment (GIVE)~\cite{agrawal_web-based_2004}.
This effect is especially strong the more common a word is (like \textit{``word''} itself), presenting more potential for false positives, and the less used an instance itself is, resulting in fewer true positives.
Initially, these false positives vastly outmatched the true, which is why several safeguards were implemented.
Matches inside other words, like \textit{engineer} in \textit{engineering}, were no longer counted, with the resulting false negatives being somewhat compensated by introduced aliases such as \textit{engineers}.
To not only match \textit{knowledge extraction}, but also \textit{extraction of knowledge} or similar, a checkbox was established to divide an instance into its containing words and check if all co-occur in a 100-character window.

None of these checks could prevent all false positives. To the authors' knowledge, automated semantic disambiguation is currently not capable of preventing them, and potentially never will be for some cases, taking earlier mentioned MIDAS
for example. 
While none of these checks could remove them entirely, they cleaned the data to a degree usable as a demonstrator of how semantically described knowledge could look like.
Every occurrence should be taken with appropriate consideration for false positives, and the presence of false positives should further support the recommendation of dedicated human oversight in the loop, preferably connected as a knowledge community.
With all that in mind, these checks, such as the co-occurrence window, allowed us to strengthen a different set of statistical rules, as described in the following subsection.  

\subsubsection{Instance-Instance-Co-occurence}
The previous paragraph introduced checks for detecting the fragment words of \textit{``knowledge engineerings''} being present within a 100 character window.
This also allows to check if any other instance co-exists in the same space, hinting at a connection between the two.
We applied the association rule mining algorithm to the identified instances across the papers and filtered the mined rules using a support value of 0.4 and lift metric \cite{brin1997dynamic} with a threshold of 1.
An example of the identified rule is the connection from the process ``knowledge management'' to the data item ``document'' with 94 \% confidence, support of 0.45, and a lift of 1.18.
Such association rules were employed to visually represent a graph, which is again too large and too dense for display here.
A preview is provided in the appendix as \secref{fig:instance_instance_co_occurrence_matrix_graph}, the entire version is in the Zenodo dataset.
Equally, false positives from the aforementioned issues, alongside co-occurrence not being a perfect proxy for relationships, resulted in only strong association rules being used for further curation.
They were added to our knowledge graph as relations between the instances, using the object properties already established for paper-instance occurrences.
All of these results culminate into the knowledge graph of
\begin{itemize}
    \item 8 classes, 9 annotation properties, 13 object properties and 957 individuals;
    \item 14556 axioms, 957 of which are class assertions, 4258 are annotation assertions and 8334 are object property assertions.
\end{itemize}
How to best access this data is described in the next subsection.

\subsection{Open Accessible Knowledge Graph}
Our knowledge graph was curated using WebProtégé\footnote{\url{https://webprotege.stanford.edu/}}~\cite{musen_protege_2015}, which is perfect for curating with a set team but not designed as an open-access knowledge infrastructure like Wikidata.
Similarly, Wikidata is designed to be a general-purpose knowledge base, not a domain-specific knowledge graph, neither for science nor aerospace engineering.
As experts like \citeauthor{sanya_challenges_2011}, \citeauthor{xie_application_2013}, \citeauthor{procko_leveraging_2022} have pointed out, however, exactly this collaborative knowledge-sharing is what the domain needs.
As such, we present our results as foundations of a future knowledge-based aerospace engineering knowledge community, on the \nameref{sec:results_orkg} and a dedicated \nameref{sec:results_wikibase} instance.

\subsubsection{Open Research Knowledge Graph\label{sec:results_orkg}}
\begin{figure}[htb!]
    \centering
    \includegraphics[clip,width=.95\linewidth, trim={0.6cm 9cm 7.6cm 2.4cm}]{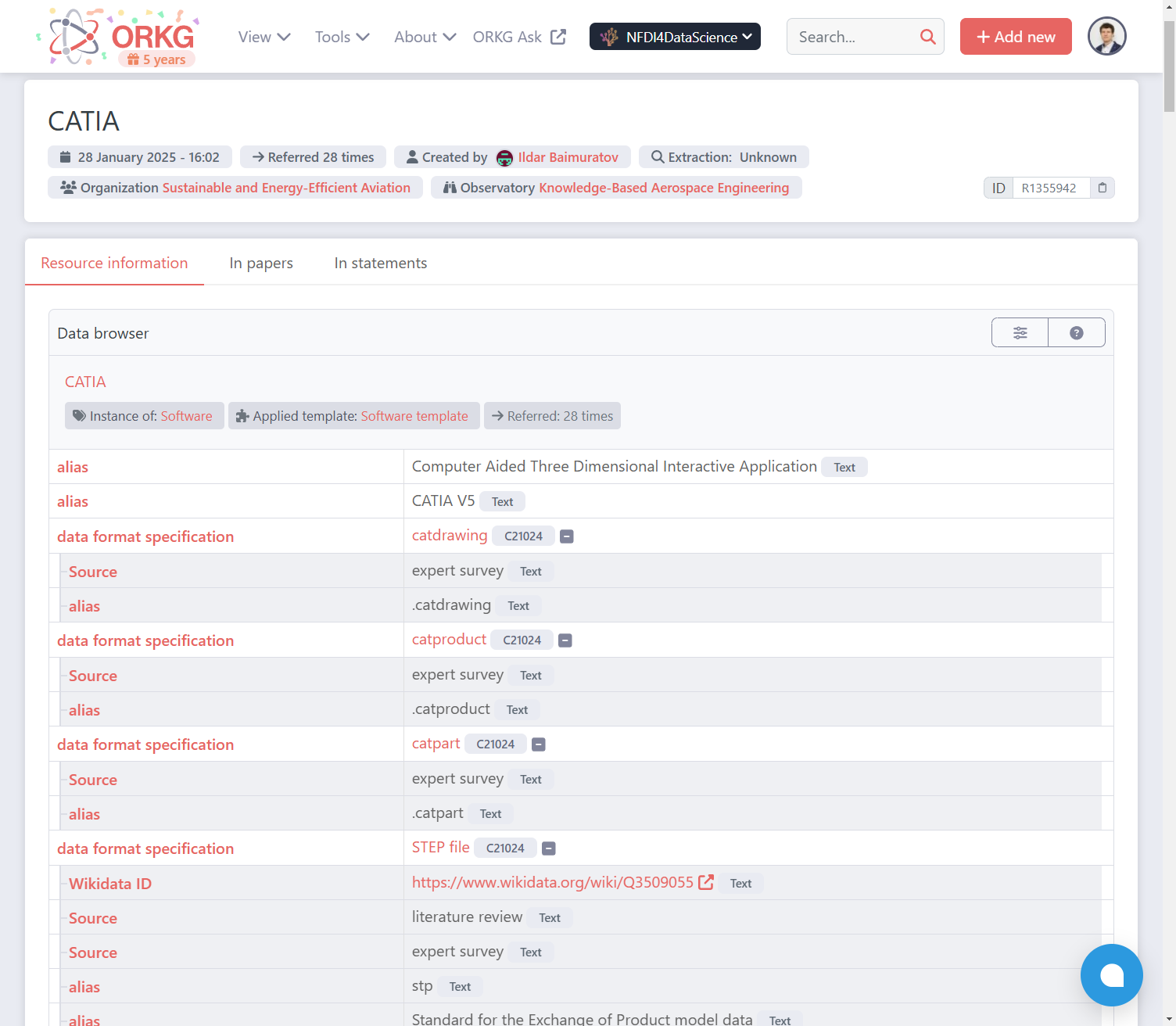}
    \caption{\gls{catia} resource information and paper occurrence view on the \gls{orkg}\protect\footnotemark{}}
    \label{fig:catia}
\end{figure}

\begin{figure}[htb!]
    \centering
    \includegraphics[clip,width=.95\linewidth, trim={0.6cm 12cm 7.6cm 8cm}]{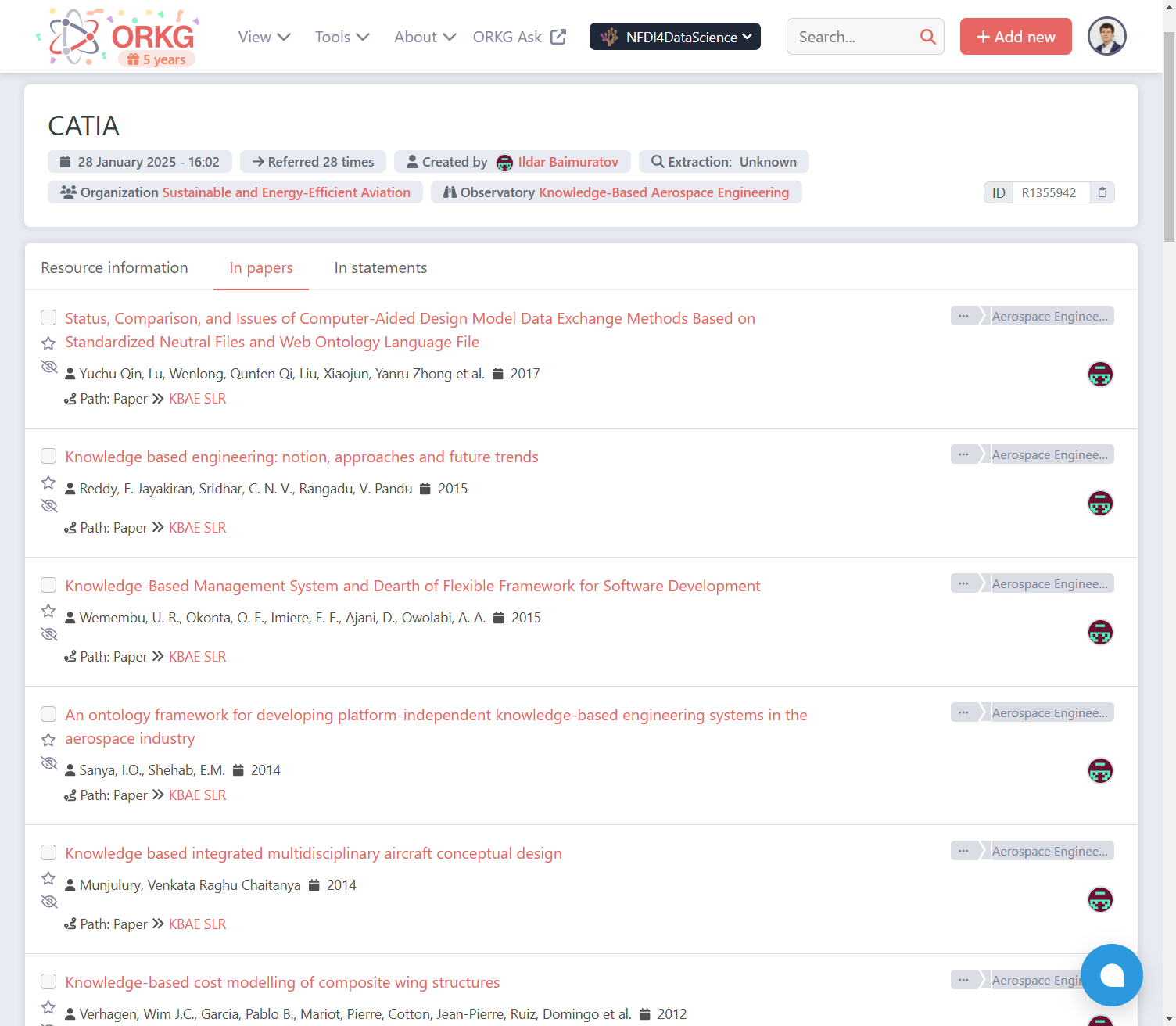}
    \caption{Occurence of the \gls{catia} resource in papers on the \gls{orkg}}
    \label{fig:catia_in_papers}
\end{figure}

\begin{figure}[bth!]
    \centering
    \includegraphics[width=.85\linewidth]{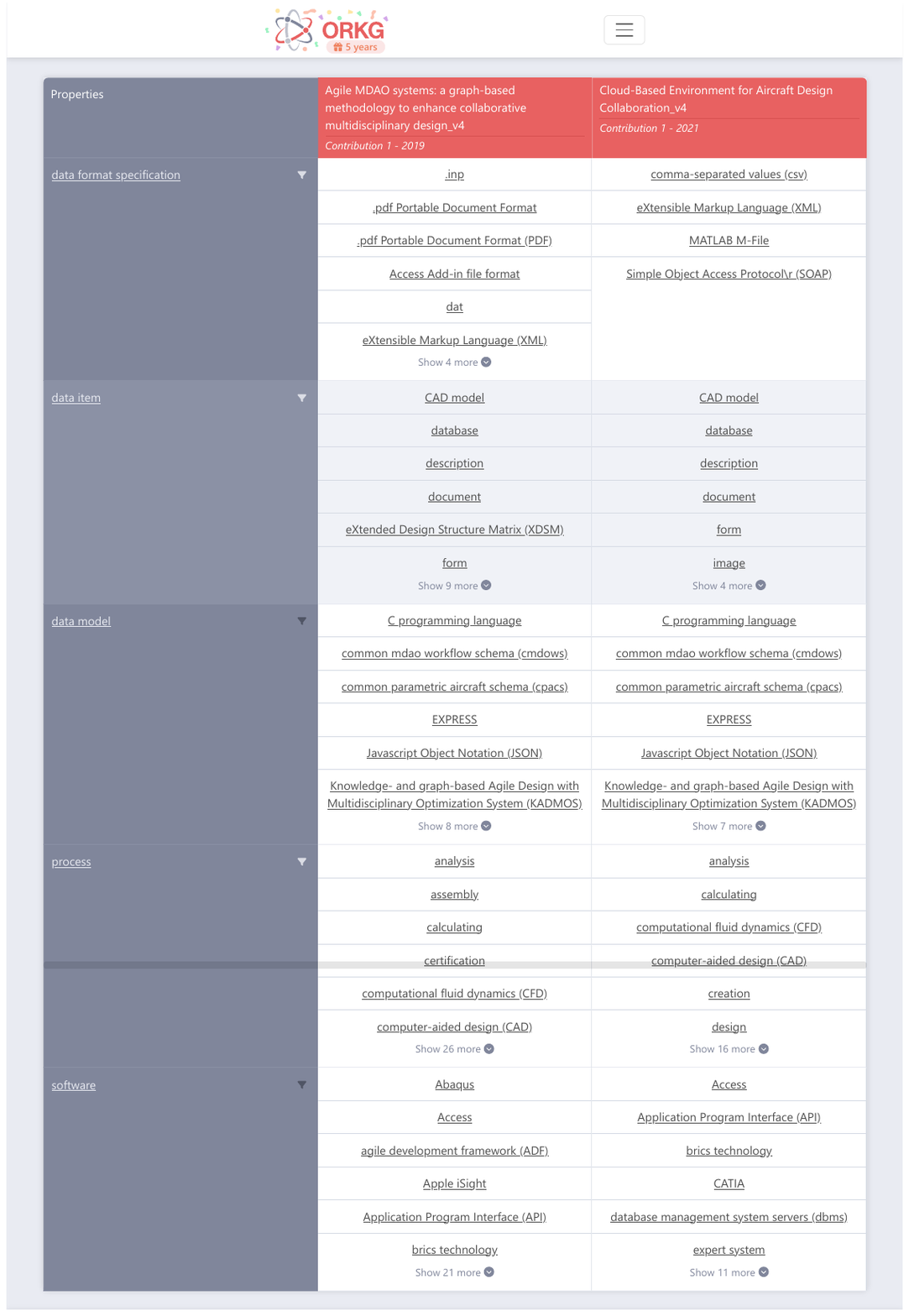}
    \caption{Comparison with filters for \gls{catia} and \gls{kadmos} applied\protect\footnotemark{}}
    \label{fig:comparison}
\end{figure}

The \gls{orkg} contains the entire knowledge graph created by this literature review:
\begin{itemize}
    \item Every single instance is an \gls{orkg} resource, providing unique identifiers for future reuse. If they are more richly described, such as \gls{catia} (\href{https://orkg.org/resource/R1355942}{R1355942})
    in \secref{fig:catia}, their relations to other resources are easily accessible.
    \item All papers are added to the \gls{orkg}, listing their contributed processes, software and data. This mapping allows identifying papers that mention a resource of interest, see \secref{fig:catia_in_papers} as an example.
    \item All reviewed papers are added to an ORKG comparison, which makes it possible to filter papers as depicted in \secref{fig:comparison}. For example, one paper listing both the \gls{kadmos} data model and the \gls{catia} software is the paper (\href{https://orkg.org/paper/R1356673}{R1356673}), representing \citet{van_gent_composing_2017}.
    The comparison is often the most desired result of a systematic literature review.
    The comparison presented here is an automatic result of properly represented paper contributions and resources in the \gls{orkg}.
    As such, any further contributions added to the knowledge graph can simply be added to the comparison, extending and updating it.
    \item The comparison is added to an ORKG review\footnote{\url{https://incubating.orkg.org/review/R819286}}. Alongside additional relevant context and navigational guidance over the \gls{orkg} artifacts, it serves as a living version of the present \gls{slr} that can be updated with new papers.
    \item Every \gls{orkg} artifact described by now is added to the ORKG observatory dedicated to Knowledge-Based Aerospace Engineering\footnote{\url{https://incubating.orkg.org/observatory/Knowledge_Based_Aerospace_Engineering}}. This observatory is specifically setup to provide a structure for a future knowledge-based aerospace engineering community. It provides a common root to every relevant node in the graph, regardless of the specific problem or context at hand. Every paper or comparison relevant to the domain can be connected to this central node, alongside visualizations, reviews and other artifacts, making it the domains' main entry point for relevant, curated knowledge.
\end{itemize}

\subsubsection{Wikibase\label{sec:results_wikibase}}
In several instances, we have highlighted the importance of Wikidata for open linked data.
A significant amount of the domain-specific software and processes did not have Wikidata items.
Changing this by steadily increasing the representation on this central knowledge graph is desirable, but not uncomplicated.
To allow as many aerospace engineers to test and experiment with semantic representations, we have setup several testing grounds:
Where the \gls{orkg} has the sandbox\footnote{\url{https://sandbox.orkg.org/}} to collect experience, we have setup a dedicated wikibase instance for the aerospace community: \href{https://aerospace.wikibase.cloud}{aerospace.wikibase.cloud}.

This Wikibase provides an example of how a sovereignly manageable, but fundamentally interconnected knowledge graph could be a central knowledge sharing infrastructure, fitted to the domain specific requirements.
Ideally, such a Wikibase is picked up by central institutions representing aerospace, such as \gls{nasa}, where MediaWiki is already used for internal knowledge management, \gls{aiaa}, \gls{dlr}, \gls{tu} Delft, or similar.
Independently, even a single engineer could populate this or their own Wikibase, or start contributing on wikidata and the \gls{orkg} right away.
The knowledge infrastructure, required by the previous authors, is available today.

Following the overview on how our identified knowledge may generally be accessed, the sections \nameref{sec:processes}, \nameref{sec:tools} and \nameref{sec:data} are dedicated to focus on class specific findings.
\pagebreak

\footnotetext{\url{https://orkg.org/resource/R1355942}}

\pagebreak

\footnotetext{\url{https://incubating.orkg.org/comparison/R819284}}

\FloatBarrier

\section{Process Findings\label{sec:processes}}
A challenge in the created ontology was to classify the obtained processes in a structured way.
As seen before, the question from \secref{sec:task7} of ``Which Aerospace Engineering \textbf{Process} is completed by which \textbf{Software} using which \textbf{Data} in which \textbf{Format} and \textbf{Schema}?'' was used to initially classify the knowledge in a selection of distinct classes.
However, to further improve the ontology structure, the processes themselves also had to be classified.
Taking the research questions as an example, we asked
``by which means, such as knowledge representations and schemas, do aerospace engineers externalize, such
as formalize and visualize, 
knowledge?''.
This externalization already categorically includes formalization and visualization, but likely other processes.
Similarly, the other questions contain
``utilize, such as organize
and interface with,''
and 
``exchange, such as transfer and
distribute, explicit  knowledge''.
These knowledge engineering processes are intertwined with engineering processes prevalent throughout all product development phases.
As such, we provide some context on the structuring of processes along the development in \secref{sec:VModelResults} and conclude this section with the findings of the \nameref{sec:process_qual} and \nameref{sec:process_quan}.

\subsection{Knowledge-Based Practices along the V-Model}\label{sec:VModelResults}
The V-model was chosen as a suitable structure for classifying the various processes found in the ontology for this work.
This involved creating processes for each of the V-model phases from Fig. \ref{fig:VModel} and thereafter classifying each process to its corresponding phase or phases.
Naturally, this classification is prone to mistakes and biases without proper support and guidance from experts for each process instance.
However, certain processes such as ``aircraft concept analysis'' can with a relatively high certainty be classified as belonging to the ``System-Level Requirements'' phase from Fig. \ref{fig:VModel}.
Processes such as ``document classification'' are on the other hand harder to classify due to their knowledge engineering nature allowing them to be valuable indifferent to the current V-model phase.

To support a proper classification of the processes, the V-model positioning part in the second survey was added as a means for receiving expert inputs on where they typically use their tasks, software, and data formats.

\begin{figure}[bt]
    \centering
    \includegraphics[width=0.70\linewidth]{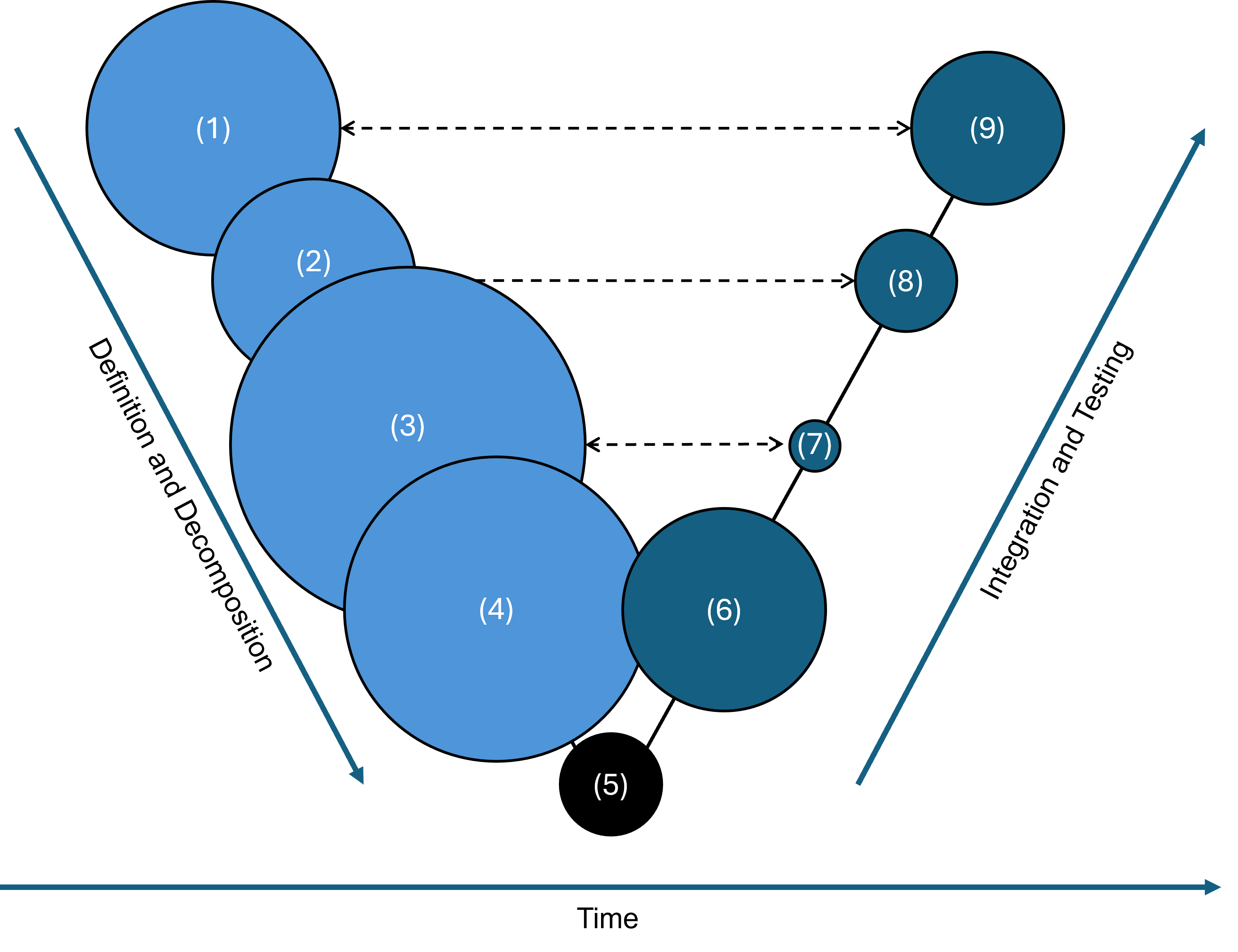}
    \caption{The answers from the second survey positioned along the V-model.}
    \label{fig:VModelResults}
\end{figure}

Even though the second survey from \secref{sec:results_survey} saw a low response rate, some initial insights on where knowledge-based practices
typically are used along the different phases of the V-model could be gained.
\secref{fig:VModelResults} illustrates the responses from the survey for where tasks related to knowledge-based practices typically are used in aerospace engineering,

The size of each circle in \secref{fig:VModelResults} corresponds to the number of entries from the positioning in each phase of the V-model according to the second survey results.
Also, the number in each circle represents the corresponding V-model phase from \secref{fig:VModel}. 
As seen, the left-hand side of the V-model is where most of the knowledge-based practices take place.
This is particularly true for the ``High-Level Design'' and ``Component Design'' phases that received the largest number of entries.
Consequently, knowledge-based practices are predominantly used in the phases of the ``Definition and Decomposition'' part of a development process and thereby indicate that many of the processes in the ontology could be classified as belonging to this side and its corresponding phases. As a result, this reduces some of the classification uncertainty for the processes that could not be classified directly from the survey answers. 
However, the positioning of survey answers along all phases of the V-model can be observed in \secref{fig:VModelResults}.
One of the explanations for this is that many engineers also deal with the corresponding phases on the ``Integration and Testing'' side.
Interestingly, there is a large difference between the subsystem requirements and the subsystem verification phases.
This may consequently indicate that engineers working on high-level design are typically less involved in the verification processes.
However, more survey responses are needed to draw a proper conclusion about this.
Overall, the results from the second survey and \secref{fig:VModelResults} indicate that these responses help with classifying the processes into the generated ontology along the V-model phases.
However, more survey participants are needed to substantiate a comprehensive and correct classification of all identified processes in the ontology. 

\subsection{Quantitative Study\label{sec:process_quan}}
In the knowledge graph, 370 instances were classified as processes.
All 164 papers mention some process,
where \citet{li_research_2018} mentions the most number of processes, 26, and \citet{adala_framework_2011} mentions the least with just three.
The most frequently mentioned process is design, 163 papers mention it, while ``flight test reporting'', ``information push'', ``semantic information transferring'', ``semantic lexicon'', ``share documents'', ``design concept formalization'', ``engineering java programming'' and ``tooling development'' were mentioned only once.
We have identified associations for 58 processes with software,
where the ``function plotting'' process is associated with the most number of software, three. Eleven processes are associated with two software instances and the remaining 46 processes are associated only with one instance.
%
There are 29 processes associated with a data item.
%
Manufacturing and reasoning are associated with five data items, while 15 processes are associated with four data items, two processes are associated with three data items, the only inference is associated with two data items, and nine processes are associated with only one data item.
%
%
Finally, 110 processes were identified with Wikidata entries, which makes almost 30\% of all processes.
%
214 processes come from literature review, 86 from expert survey, and 58 from curation, where 6 processes have more than one source.

\subsection{Qualitative Study\label{sec:process_qual}}
The main process types of interest were the aforementioned externalization, utilization and exchange of knowledge.
As such, any processes such as formalization, visualization, interfacing and distribution were sought after.
Evidently, various methods for formalization were found, as will be further described in \secref{sec:data}.
Different pathways of interfacing were also described, ranging from user-centered interfaces like Excel or a \gls{sysml} diagram described in \citet{herrero_system_2022} to highly decentralized and automatized tool-chains, as discussed by \citet{flink_orchestrating_2022} and many others.
These tools attempt to address the issues highlighted for over a decade, e.g. by \citet{jackson_progress_2006} when stating that ``The current state-of-the-art is such that several staff-months if not staff-years are required to 'rehost' each release of a flight dynamics model from one simulation environment to another one''.
These tools will be further discussed in \secref{sec:tools}.

Especially troubling, however, were the findings towards distribution and visualization.
The general findings already described the issue of distributing knowledge in aerospace.
\citet{royal_advancing_2022} specifically states that aircraft development in a ``central hub environment is generally not a practical option'', since partners ``wish – or are even forced'' to keep their data and tools on their own premises.
Yet, \citet{vucinic_multidisciplinary_2010} emphasizes the importance of expert synergy, particularly between computer scientists and engineers, focussing on the necessity of interpersonal knowledge exchange through means of visualization.
``Visualization'', however, ``is only an issue for human interpretation'', as \cite{bohnke_data_2009} state. 
In a field where sharing knowledge with others is already a low, potentially negative priority, establishing a common language, particularly a visual one, is tricky. 
As such, we have identified a knowledge visualization gap, to be discussed in the following subsection.

\subsection{Gap Analysis}
As established, aerospace engineering knowledge is complex, highly heterogeneously modeled, scarcely exchanged between humans and even less between organizations.
As such, visualization is of very little priority, barely not existent beyond several classes of diagrams.
Of particular interest is that the forefront of knowledge engineering, \gls{owl},
lack of comprehensive visualization means at scale.
WebVOWL\footnote{\url{https://github.com/VisualDataWeb/WebVOWL}} for example seems not capable of displaying instance nodes.
This functionality is only available in tools like OWLviz\footnote{\url{https://protegewiki.stanford.edu/wiki/OWLViz}}, an extension for the desktop Protégé, and WebProtégé\footnote{\url{https://webprotege.stanford.edu/}}, the more modern, lightweight web version, which was primarily used for in this work.
To the authors' knowledge, both are state-of-the-art and, while they are to be commended for their capabilities, are still beneath today's requirements.
WebProtégé has the cleanest display and untangling of resources, but little customization beyond that, becoming impractical beyond a few dozen connections.
OWLviz provides exactly this customization, but its default sorting are practically unusable at scale, leading to hours of manual time investment required for displaying a single graph, only for this time investment to become largely irrelevant as soon as the graph is updated and loading an old save is no longer meaningful. 
Since not only aerospace, but many other domains need to describe ever more complex systems at ever more degree of detail, it is an alarming signal that today's state of the art appears to be a decade old.
An updated, more flexible and better way of visually representing knowledge in ontologies is therefore a desirable feature and gap yet to be filled.

\section{Software Findings\label{sec:tools}}
By which means, such as knowledge bases and services,
do aerospace engineers
utilize, such as organize and interface with,
explicit knowledge?
By which means, such as tools and methods,
do aerospace engineers
exchange, such as transfer and distribute,
explicit knowledge?
These questions particularly lead the inquiry in our survey.
In short, which tools do aerospace engineers use to externalize, utilize, and exchange knowledge?
As with the previous and following section, the findings regarding this question are separated into \nameref{sec:software_quan} and \nameref{sec:software_qual}, concluded by the \nameref{sec:software_gap}. 

\subsection{Quantitative Study\label{sec:software_quan}}
The 182 tools were identified,
where 127 come from the literature review, 49 from expert surveys and only 1 from curation.
%
The most number of tools mentioned, 16, is in \citet{herbst_development_2018}.
``Access'' was mentioned in 149 papers, followed by ``Word'' mentioned with 102 mentions and API with 74, while 14 tools were mentioned only once
The first two are pristine examples of false positives since both ``access'' and ``word'' are common lexemes.
There are 29 associations between software tools and processes.
%
Software associated with the most number of processes is logically Access, with 31 connections, followed by MATLAB with ten, and both Simulink and Wolfram Mathematica with four processes, while 17 software instances are associated only with one process.
%
Four tools are associated with a data item,
%
with ``Word'' having the most number of data items, eight.
%
Eight tools are associated with a data format,
%
where TAU has the most number of connections, eight, and three tools a associated with only one format.
%
Finally, five tools are associated with a data model.
%
%
When trying to disambiguate tools using Wikidata, only 37\% of the initial 135 software instances were linked.
%

\subsection{Qualitative Study\label{sec:software_qual}}
Tools generally can be classified into one of the following two groups:
Highly popular tools, such as CATIA, MATLAB or Excel, are universally usable and accessible, and niche specialized tools are mostly used by their developers, like cpacspy or DAVEtools. 
Especially the latter group is prone to be entirely unavailable, discontinued, or rebranded by the time we identified them, such as Vivisimo, now called IBM Watson Explorer for example.
Generally, the popular group was highly condensed, and the lower popularity was widely distributed. 

In trying to disambiguate these resources using Wikidata and web search, we coincidentally came across a r/AerospaceEngineering Reddit post\footnote{\url{https://www.reddit.com/r/AerospaceEngineering/comments/mr9unc/what_software_do_you_use_as_an_aerospace_engineer/}}, largely confirming this finding of a very small set of software being used for a majority of all tasks.
As for more scientific sources, some publications described workflows throughout interoperable software, called software workflow management frameworks, such as the \acrfull{rce}\glsunset{rce}, particularly for coupling \gls{mdao} processes.
In these large conceptual mappings, several tools were just abbreviations, which sometimes simply referred to steps, not tools, e.g. ``TR'' for ``Term Recognition''.
Some of these abbreviations could be disambiguated, like CoOM referring to a tool developed at \gls{tu} Darmstadt, or were difficult to find in the literature, such as COCO, a boundary layer solver developed in 1998 at \gls{tu} Braunschweig \cite{coco_radespiel-1998}.
Especially unhelpful for identification is the prevalence of naming software after existing concepts, such as Fiji or SALOME, which could only be mapped if the tool in question was described well enough.
While SALOME even had a Wikidata entry, the majority were barely described at all, with some publications not even mentioning the tools they used, or just mentioning their concepts like ``Python application''.
For example, a process called \gls{pido} (\href{https://www.wikidata.org/wiki/Q7119347}{Q7119347})
will have \gls{pido} platforms and applications associated with it, but these might not be named.

Concluding this overview of what software is used, we can say with relative certainty that the answer is generally ``Microsoft Office, a small set of widely applicable scripting tools, one or two use-case specific software to solve exactly one given problem, and a self-developed software with short development lifespan that has little usage outside the current project''.

\subsection{Gap Analysis\label{sec:software_gap}}
As for our leading questions, which tools do aerospace engineers use to externalize, utilize, and exchange explicit knowledge, the general answer appears to be ``barely any, if at all''.
Knowledge is usually not formalized, but distributed via resource management on the human resources level, in reports, meetings, and direct communication.
Explicit knowledge bases are widely mentioned, recommended, established, then rarely used, abandoned, and eventually cease to exist.
If they continue to exist, they are generally closed to the outside, such as at \gls{nasa} or Siemens.

To our knowledge, there are no open-access aerospace knowledge graphs, wikibase instances, \gls{orkg} observatories, or similar.
The closest thing could be the Wikipedia Aerospace Engineering glossary\footnote{\url{https://en.wikipedia.org/wiki/Glossary_of_aerospace_engineering}}, all other open and collaborative knowledge bases seem to have been discontinued.
As such, it is vital that collaborative infrastructure is reused, such as a wikibase.cloud instance, to prevent future knowledge graphs from fading like every previous iteration apparently has.

\section{Data Findings\label{sec:data}}
In this section, we present our findings related to RQ1: By which means, such as knowledge representations and schemas, do aerospace engineers externalize, such as formalize and visualize, knowledge? Here, we discuss the items classified in our knowledge graph as data models, data items, and data formats.

\subsection{Quantitative Study}
Totally, 89 data formats were mentioned in
46 papers,
%
%
47 coming from the literature review and 37 from the surveys.
%
The most commonly mentioned data formats are PDF (32 mentions), VRML (10) and RDS (6).
We identified Wikidata URIs for 61 (69\%) of the formats.
Most knowledge not exchanged in PDF or Office format is based on one of these two.

Regarding data items, 81 instances were identified
over all papers and
44 of them (54\%) have Wikidata URIs.
%
%
%
In a single paper, the number of data item mentions varies from one (two papers) to 14 \cite{xie_application_2013}.
%
The most common data items are description (144 mentions), database (138), and material (108).
Description, form, document, and database have 21, 20, 19, and 17 associated processes respectively, while eight data items are associated only with one process each.
Regarding sources, 62 data items come from the literature review and 14 from curation.

Finally, we identified 70 data models
mentioned over 142 papers,
63 of them come from the literature review, four from the expert surveys, and one from curation.
%
The number of data model mentions in a paper ranges from one (64 papers) to seven \cite{procko_leveraging_2022}.
%
The three most common data models are EXPRESS (117 mentions), \gls{uml} (52) and \gls{sparql} (26) with seven data models mentioned only once.
Three data models, EXPRESS, XML, and IDEF1X are associated with a software instance, where EXPRESS has the most number of connections, three.
%
Only 39 data models were identified with Wikidata URIs, which makes 56\%.
%

\subsection{Qualitative Study}

Knowledge-based aerospace engineering employs various data models, languages, and ontologies. Qualitative study showed, \gls{xml} is the most commonly used modeling language, as described in \secref{tab:data_models}. Other widely adopted options include \gls{uml} or \gls{sysml}, \gls{rdf} and \gls{owl} sometimes enhanced with \gls{swrl}. Studies utilizing semantic modeling with \gls{owl} predominantly rely on the Protégé tool. Additionally, a few works make use of MS Access \cite{haney_data_2016} and MATLAB \cite{herbst_development_2018}.

\begin{table}[thb]
    \centering
    \caption{Languages used for data modeling}
    \label{tab:data_models}
    \begin{tabular}{cc}
        \textbf{Language} & \textbf{Papers} \\
        \hline
        XML & \cite{vargas-hernandez_development_2003} \cite{jackson_progress_2006} \cite{agrawal_web-based_2004} \cite{nagel_communication_2012} \cite{van_gent_agile_2019}  \cite{munjulury_knowledge_2014} \cite{pulvermacher2000space} \\
        UML & \cite{lu_data_2005} \cite{sung2007component} \cite{berquand_text_2021} \cite{van2008knowledge} \cite{oh2001mapping} \\
        SysML & \cite{de_leon_intelligent_2021} \cite{bohnke_data_2009} \\
        RDF & \cite{dadzie_applying_2009} \cite{khilwani_managing_2016} \cite{herrero_system_2022} \cite{bang_daphne_2018} \cite{padilha_enabling_2023} \\
        OWL & \cite{zhao_application_2020} \cite{curran_knomad_2010} \cite{franzen_ontology-assisted_nodate} \cite{kuofie_radex_2010} \\
        SWRL & \cite{hoogreef_multidisciplinary_2015} \cite{roelofs_automatically_2021} \cite{markusheska_implementing_2022} \\
    \end{tabular}
\end{table}

Moreover, researchers develop specific data models implemented with the languages mentioned above. These include bond graphs \cite{vargas-hernandez_development_2003}, \gls{mdo} Framework for Analysis and Design (\gls{mdo} FAD) \cite{agrawal_web-based_2004}, \gls{cpacs} \cite{nagel_communication_2012} and \gls{cmdows} \cite{van_gent_agile_2019} represented with XML; \acrfull{dtad} \cite{herrero_system_2022} and Codex \cite{padilha_enabling_2023} represented with \gls{rdf}; \gls{step}\footnote{\url{https://www.iso.org/obp/ui/\#iso:std:iso:10303:-1:ed-3:v1:en}} formulated with the EXPRESS language; \gls{kadmos} \cite{van_gent_agile_2019} and \acrfull{mdax} \cite{ap_mdax_2020} implemented in Python; \acrfull{radex} \cite{kuofie_radex_2010} modeled with \gls{owl}; \citet{ezhilarasu2021development} develop \acrfull{anfis}. Additionally, well-established data models are used. Thus,  \citet{roelofs_automatically_2021} utilize \acrfull{bfo} and WordNet is utilized in \cite{faris_semantic_nodate} and \cite{gargiulo2014aerospace}.

Ontologies related to the aerospace domain include the NASA Taxonomy\footnote{\url{https://github.com/nasa/dictionaries}}, XCALIBR \cite{putten2008ontology}, the Aerospace Ontology \cite{malin2007basic}, \acrfull{atmonto} \cite{keller2017nasa}, Airspace System Ontology \cite{miller2017ontology}, Avionics Analytics Ontology \cite{insaurralde2018uncertainty}, Orbital Space Environment Domain Ontology \cite{rovetto2017ontology}, Orbital Debris Ontology \cite{rovetto_orbital_2020}, and Space Situational Awareness Ontology \cite{rovetto_preliminaries_2016}. Additionally, ontological modeling without specific naming is described in \cite{halvorson_ontology_2022}, \cite{wang_knowledge_2018}, \cite{verhagen_ontological_2011}, \cite{neumayr2017semantic}, \cite{okoh2014development}, \cite{wright2008semantic}, \cite{wicks2011ontological}, and \cite{cambresy2010ontology}.

\subsection{Gap Analysis}

\paragraph{Standards}
For interoperability, knowledge-based aerospace engineering requires a common language. While we were able to identify 70 data models in the literature for our knowledge graph, we could not identify a single one that is widely adapted throughout the domain.
Even capable models rarely find applications outside their originator's publications, like \gls{dlr} for \gls{cpacs} or \gls{tu} Delft for \gls{cmdows}.
This is further substantiated by recent works such as \citeauthor{tikayat_ray_development_2024}~\cite{tikayat_ray_development_2024}, creating a glossary from the scarce aerospace resources publicly available.
While this work aims to unify aerospace terminology openly available and share its software just as such, the glossary itself is once again unavailable.
All previous attempts are either no longer available or never were due to organizational boundaries.
While tailored knowledge is highly proprietary, a general ontology does not conflict with system boundaries.
Even proprietary knowledge can still be available ``open by default'' within these boundaries, as is already the default in organizations such as \gls{nasa} and Siemens.
If standards are established by these organizations, they will facilitate interoperability and reusability, especially if knowledge facilitators such as the \gls{aiaa} or the \gls{omg} support them.

\paragraph{Semantics}
While having a shared glossary is commendable, truly interoperable knowledge engineering requires crafting a terminology by hierarchical structuring these terms, an ontology by separating entities into concepts and individuals, and a connected knowledge graph using defined properties. 
A higher level of interpretability is supported by systems utilizing \gls{rdf}, \gls{owl}, and especially \gls{swrl} language.
However, XML-based models, such as \gls{cpacs}, are prevalent in the aerospace domain.
Though most of them ensure interoperability, they provide primarily human-, not machine-interpretability.
This is a common issue throughout aerospace software and data, being mainly used to exchange knowledge between individuals, not larger, machine-supported systems.

\section{Discussion and Future Work\label{sec:discussion}}
We will briefly discuss general aspects regarding \nameref{sec:disc_kbeae}, rounded up with \nameref{sec:disc_lr} in the domain.

\subsection{Knowledge-Based Aerospace Engineering\label{sec:disc_kbeae}}
\subsubsection{Aerospace Knowledge Engineers}
We identified a divide between Aerospace Engineers and Knowledge Engineers.
While both collaborate towards Knowledge-Based Aerospace Engineering, their goals are not necessarily aligned:
One primarily solves a current issue, the other is focused on aiding the solution of future instances of this current issue.
The first prefers a practical solution, and the latter a theoretically interoperable one.
While many works have highlighted the difficulty of approaching the aerospace domain from a knowledge management standpoint, a better approach seems to be to fundamentally think of both of them in a coupled system.
If aerospace engineers work in a team with dedicated knowledge engineers, the domain engineers can focus on the current task, while the knowledge engineers interfaces to previous and future use cases.
There are several works that already suggested or implemented this, indicating this to be a very promising composition for future Knowledge-Based Aerospace Engineers.

\subsubsection{Knowledge-Based Practices and the V-Model} 
The study of knowledge-based practices positioned along the V-model model phases from \secref{sec:processes} gave some initial insight into where tasks, software, and tools are used.
However, more responses to the survey are really needed to draw proper conclusions on this particular part of the presented study. 
Nevertheless, the survey shows that the classification of processes in the created ontology from the literature survey may be supported by such survey responses.

Also, the survey was distributed to a variety of aerospace engineers but not all might have an understanding of what knowledge-based practices imply. Consequently, a future iteration of the survey would include more descriptions and examples of this. This would then add to more accurate responses to the survey and corresponding entries in terms of positioning along the V-model phases.
The next survey would also be targeted at a wider group of aerospace engineers in order to gain more diverse input on knowledge-based practices. 

The classification of the processes in the ontology was done manually for this work. However, this can be a challenging task due to the inherent complexity and ambiguity of knowledge-based practices, especially when expert input is not available. Without domain expertise, there is a risk of misclassification or oversimplification, which could lead to inaccuracies in the ontology representation. 
This issue becomes even more pronounced as additional future survey responses may be integrated, increasing the diversity and volume of entries that must be accurately categorized. 
A possible future venture would be to investigate the use of, for example, a Large Language Model (LLM) as an effort to automize the classification process and thereby add to the overall scalability of the study. This could also enable quick additional classifications according to other frameworks than the V-model in the same ontology. Such a study could potentially add further insights into knowledge-based aerospace engineering practices in general and thereby contribute supplementary perspectives to the answers to our research questions.

\subsection{Perspectives on Literature Reviews\label{sec:disc_lr}}
\subsubsection{Lessons Learned}
This work is based on the \gls{swarm-slr} and tested the applicability of several approaches to facilitate a knowledge-based aerospace engineering literature review.
We sourced 245 relevant documents out of a set of over 4000 using the established workflow, read over 150 full papers, dissertations, and articles to craft a preliminary ontology, and statistically evaluated all 1000 available PDFs.
Aligning data models and terminology even among the authors of this work was a significant endeavor, but eventually resulted in several lessons learned:
The \gls{swarm-slr} workflow~\cite{wittenborg_swarm-slr_2024} is sufficiently capable until including Task 5. 
Starting with Task 6, specifically with requirement 51. \textit{annotate collaboratively}, several lessons were learned:
\begin{enumerate}
    \item It is highly advised to \textbf{setup a WebProtégé instance}
    and spend a few minutes familiarizing all involved researchers with the software and establishing simple conventions.\\
    Initially appealing approaches, such as annotating inside a PDF reader, in Obsidian, Zotero, a CSV, or even Google Sheets, are no replacement for the semantic disambiguation required later on, including entity linking, synonym, and polynoym handling, etc.    
    The benefits of proper up-front knowledge management are obvious even to this work, where having started with this work could have easily saved dozens of hours of untangling 700+ instances.
    \item \textbf{Integrating Wikidata as early as possible} is strongly recommended.\\
    Wikidata is an invaluable source of structured information and helps to disambiguate concepts and individuals.
    Even initiating an instance such as \href{https://aerospace.wikibase.cloud}{aerospace.wikibase.cloud} could be beneficial, setting a starting point for future collaboration while alleviating the pressure of interacting with an established knowledge community.
    \item \textbf{Expert surveys are not efficient semantic data sources}, but they are helpful to differentiate between the sanitized environment of a publication and everyday problems.\\
    Responses require large amounts of manual post-processing and provide the participants little in return, leading to low response rate and immediate drop-off after initial distribution.
    A dynamic approach like collaborative editing in a Wikibase or the \gls{orkg} might be much more rewarding.
\end{enumerate}

\subsubsection{Reuse and Extend This Work}
While this work provides substantial work towards semantifying an interoperable aerospace engineering knowledge graph, it is still just a first step.
This initial momentum should lower the barrier to entry for the following endeavors:
\begin{enumerate}
    \item \textbf{Establish a common taxonomy in your organization.}  Taking this work and examples like the NASA taxonomy as a starting point, this step should be straightforward to implement.
    \item \textbf{Annotate your contributions.} It is very likely that valuable knowledge was missed in this literature review, ranging from unavailable documents to simply unidentified articles outside the search scope. To ensure your contributions partake in future overviews, constructing their \gls{fair} representations, ideally already at inception, is of high importance.
    \item \textbf{Invest in manual \glsentrylong{km}}, since as highlighted by many authors, an equal, if not larger amount of work is done down the line either way when future work attempts to pick up where the current left of. Even individual engineers may save time by iteratively building upon and reusing their own knowledge base. This will also address the issue of accuracy highlighted in this work, not only clearing up false positives but also providing previously missed true positives.
\end{enumerate}

We also encourage systemic change:
\begin{enumerate}
    \item \textbf{Establish a common standard for aerospace engineering data models}. Several candidates are available, none are currently perfect, but likely, some are sufficient.
    \item \textbf{Take up centralized knowledge curation} in the aerospace Wikibase or the \gls{orkg} observatory.
    Since this literature review has surfaced the phrasing ``XY was a database for aerospace engineering knowledge'' or similar several times, it is highly recommended to steer close to established giants like Wikidata before creating another knowledge base and waiting for the corresponding project funds to dry out.
\end{enumerate}

Despite the wealth of results, findings, and recommendations, this work has only scratched the surface of what is still required in knowledge-based aerospace engineering.
Our ontology is hundreds of entries long, but facing the scale required to adequately represent the domain, it is merely a first step, requiring expansion in width, depth, and interconnectivity.
Our paper representations in the \gls{orkg} are even greater in number, but they lack the hours of curation invested in the items they display and require individuals to collaboratively scale the literature flood.
This review concludes with an extensive, but inherently limited glance on which knowledge aerospace engineering can rely, but it is up to the domain to build upon the provided foundation.

\section{Conclusion\label{sec:conclusion}}
In this \acrfull{slr}, we focused ourselves on \acrfull{kbe} practices in the aerospace domain. Compared to related reviews, ours features a software-assisted methodology and follows \gls{fair} principles, providing reusable and machine-readable artifacts. Namely, our results include a Zotero library, a Zenodo repository, and a knowledge graph with identified papers, terms, and relations between them. The latter is also represented in the \acrfull{orkg} for open public access. Contentwise, our main contribution is identifying through literature review, expert surveys, and manual curation more than 700 specific terms, or items, related to aerospace knowledge-based engineering. We classified them into three main categories: processes, software, and data. For papers and items of each category, we conducted both qualitative and quantitative studies, discovering gaps in stat-of-the-art.

This systematic literature review on knowledge-based aerospace engineering highlights the evolving landscape of how explicit knowledge is formalized, utilized, and exchanged in this highly complex domain. 
Despite the aerospace industry's longstanding tradition of applying \gls{kbe} methodologies, significant gaps in interoperability, standardization, and adoption still persist. 
These facts, in connection with deficiencies in compliance with the \gls{fair} principles and in addition to data protection topics and the necessary efforts for creating and maintaining \gls{kbe} systems hamper a wide and standardized application of this approach.
While the present systematic literature review uncovered numerous suitable tools and frameworks, such as \gls{cmdows} and \gls{cpacs}, their widespread adoption remains limited. 
Challenges in cross-disciplinary collaboration, proprietary knowledge sharing, and a lack of unified data standards hinder the full realization of KBE's potential. 
Moreover, the divide between domain-specific engineers and knowledge engineers further stresses the need for collaborative frameworks that bridge practical and theoretical knowledge applications.
Our findings demonstrate the urgent need for standardized, interoperable ontologies tailored to aerospace engineering. 
Establishing such standards can drive innovation, streamline workflows, and facilitate sustainability goals. 
Future efforts should focus on fostering open-access knowledge repositories, enhancing semantic data models, and integrating knowledge engineering practices more effectively into engineering workflows.

In conclusion, this review sets a precedent for structured, semantic-based approaches to managing aerospace engineering knowledge. 
By advancing these principles, research, and industry can achieve more efficient design processes, enhanced collaboration, and a stronger commitment to sustainable aviation.

\section*{Appendix}

\begin{multicols}{3}[\captionof{lstlisting}{{\textbf{RQ2:} By which means, such as knowledge bases and services, do aerospace engineers utilize, such as organize and interface with, explicit knowledge?}}]
    \begin{lstlisting}[
    label=col:questions2,
    basicstyle=\footnotesize,
    ]
    ### knowledge base
    knowledge base:: 10
    knowledge representation:: 8
    repository:: 8
    structure:: 6
    expert system:: 6
    software:: 5
    service:: 5
    system:: 5
    method:: 5
    archive:: 5
    catalog:: 5
    knowledge:: 5
    knowledge management:: 8
    tool:: 3
    standard:: 3
    information:: 3
    network:: 3
    format:: 2
    wiki:: 2
    data:: 1
    platform:: 1
    
    ### utilize
    utilize:: 10
    interface:: 10
    manage:: 8
    organize:: 8
    access:: 7
    integrate:: 7
    leverage:: 6
    analyze:: 6
    process:: 6
    query:: 4
    
    ### Formats
    API:: 5
    protege:: 3
    Neo4J:: 3
    XML:: 2
    JSON:: 2
    RDF:: 2
    OWL:: 2
    \end{lstlisting}
\end{multicols}

\begin{lstlisting}[
    label=col:query2,
    basicstyle=\footnotesize,
    caption={\textbf{RQ2:} Query with 806 Google Scholar results}
    ]
    "aerospace engineering" OR "aviation engineering" "knowledge base" OR 
    "knowledge representation" OR "knowledge repository" OR "knowledge structure" utilize OR
    interface OR manage OR organize OR access OR integrate API OR prot\'eg\'e OR Neo4J OR JSON
\end{lstlisting}

\begin{multicols}{3}
    [\captionof{lstlisting}{{\textbf{RQ3:} By which means, such as tools and methods, do aerospace engineers exchange, such as transfer and distribute, explicit knowledge?}}]
    \begin{lstlisting}[
    label=col:questions3,
    basicstyle=\footnotesize
    ]    
    ### knowledge representation
    knowledge representation:: 10
    Knowledge graph:: 10
    Ontology:: 10
    information representation:: 8
    Taxonomy:: 8
    Semantic Web:: 8
    knowledge engineering:: 7
    schema:: 7
    expert system:: 6
    knowledge:: 5
    data representation:: 4
    semantic:: 4
    information:: 3
    repository:: 3
    documentation:: 3
    graphic:: 3
    index:: 3
    formal:: 2
    standard:: 2
    format:: 2
    structure:: 2
    graph:: 2
    plot:: 2
    mapping:: 2
    wiki:: 2
    technique:: 2
    classification:: 2
    categorization:: 2
    data:: 1
    map:: 1
    computer-aided design:: 1
    CAD:: 1
    computer-aided design:: 1
    virtual reality:: 1
    VR:: 1
    simulation:: 1
    platform:: 1
    report:: 1
        
    ### externalize
    externalize:: 10
    represent:: 10
    formalize:: 8
    visualize:: 8
    store:: 4
    structured:: 4
    articulate:: 3
    express:: 3
    manifest:: 3
    embody:: 2
    model:: 2
    document:: 1
    codify:: 1
    transfer:: 2
    explicit knowledge:: 8
    
    ### Formats
    RDF:: 4
    OWL:: 4
    XML:: 4
    CPACS:: 4
    SysML:: 4
    UML:: 4
    JSON-LD:: 3
    JSON:: 2
    YAML:: 2
    GraphML:: 1
    XSD:: 4
    \end{lstlisting}
\end{multicols}

\begin{lstlisting}[
    label=col:query3,
    basicstyle=\footnotesize,
    caption={\textbf{RQ3:} Query with 817 Google Scholar results}
    ]
    "aerospace engineering" OR "aviation engineering" "explicit knowledge" OR
    "knowledge dissemination" transfer OR share OR exchange discipline OR domain
\end{lstlisting}

\begin{figure}
    \centering
    \includegraphics[width=1\linewidth]{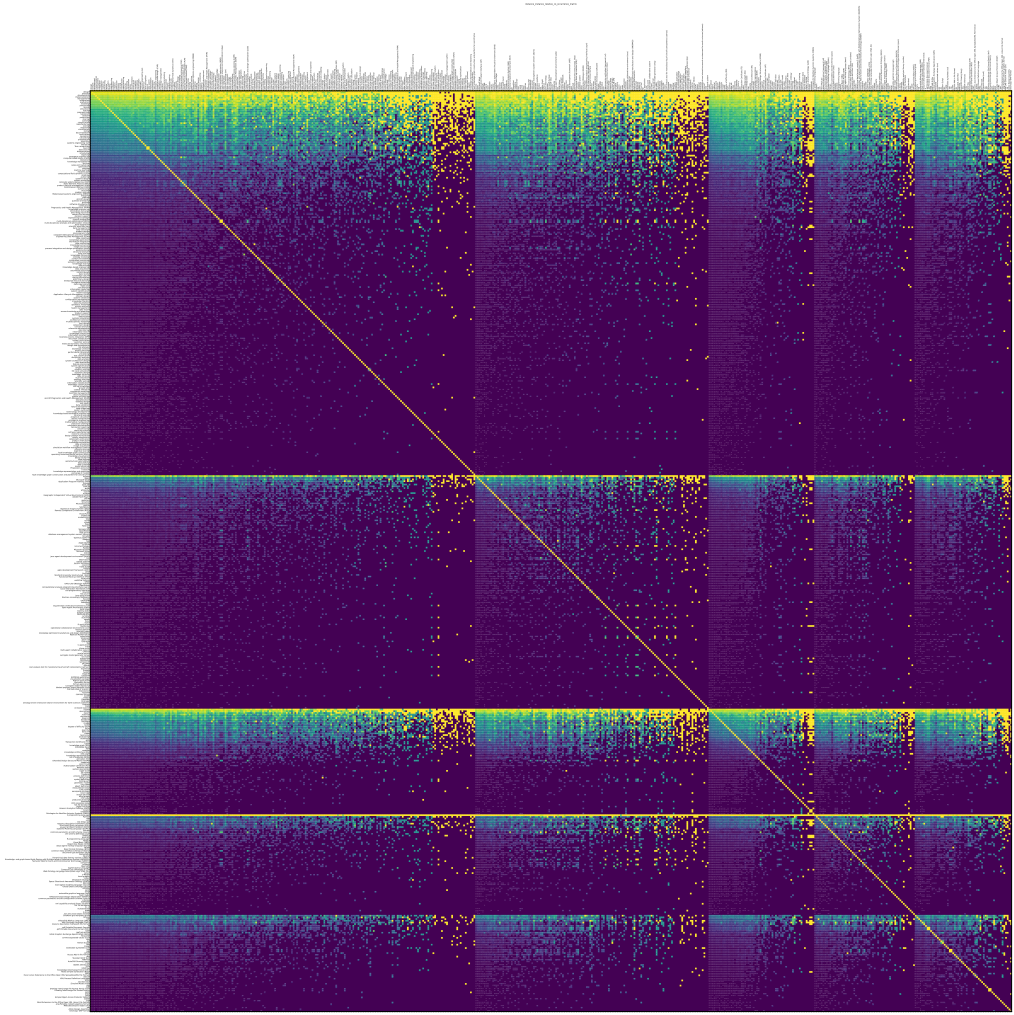}
    \caption{Instance-instance relative-co-occurrence Matrix. From purple (0) to yellow (1), each cell represents the likelihood of a given item (column) occurring within 100 characters in the same document as a given other item (row). Instances are Sorted as processes, software, data items, data models, data formats, and within that by total occurrence. The diagonal is 1, since the rows and columns are identical, and every word co-occurs with itself 100 \% of the time. The horizontal yellow stripes indicate that an item is so common in every document that it practically co-occurs with everything. The further down a yellow cell is (within its category block), the more meaningful it is, since it means a rare instance co-appears with another rare instance. Equally, if Both $i,j$ and $j,i$ are yellow, both items $i$ and $j$ occur when the other does, indicating a bi-directional, stronger relationship.}
    \label{fig:instance_instance_relative_co_occurrence_matrix}
\end{figure}

\begin{figure}
    \centering
    \includegraphics[width=1\linewidth]{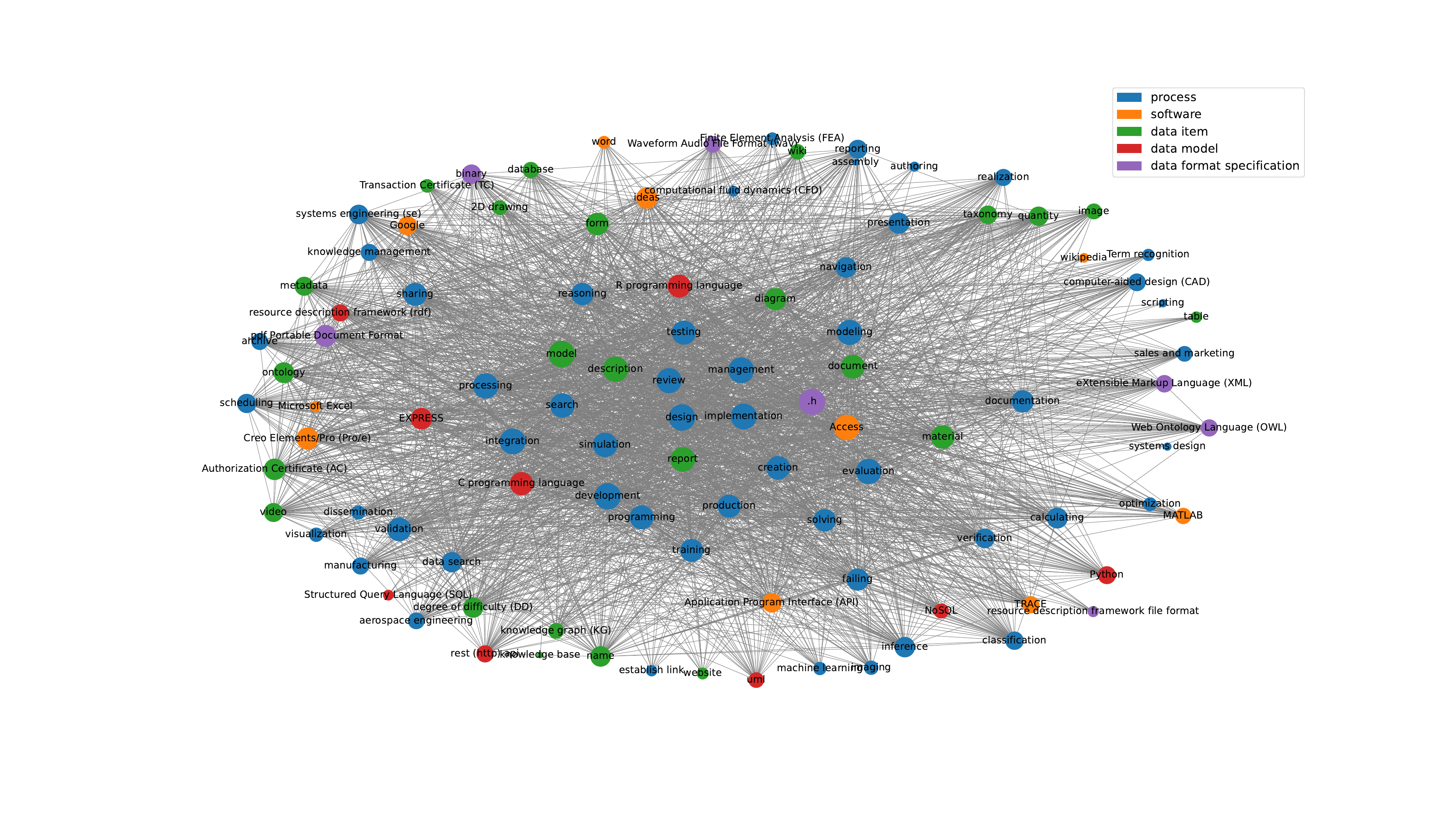}
    \caption{Instance-instance co-occurrence Graph. Taking after \secref{fig:instance_instance_relative_co_occurrence_matrix}, it represents the strongest relationships between instances, color-coded by their class and sized according to their occurrence. It was primarily an indicator of false positives, where it is visible that software like ``UG'', an acronym identified in the literature, but still unresolved to its meaning, is far too prevalent. Several other issues could be identified and resolved using this intermediate validation method, but as evident, some others remain.}
    \label{fig:instance_instance_co_occurrence_matrix_graph}
\end{figure}

\section*{Use of AI tools declaration}
During the preparation of this work the author(s) used Grammarly, LanguageTool, DeepL, and ChatGPT in order to check grammar, refine language, summarize and improve clarity. After using this tool/service, the author(s) reviewed and edited the content as needed and take(s) full responsibility for the content of the publication.

\section*{Funding Sources}

The present study was co-funded by the Deutsche Forschungsgemeinschaft (DFG, German Research Foundation) under
Germany’s Excellence Strategy EXC 2163/1 Sustainable and Energy
Efficient Aviation Project ID 390881007, as well as the NFDI4Ing project funded by the German Research Foundation (project number 442146713) and NFDI4DataScience (project number 460234259).

\section*{Acknowledgments}
We would like to acknowledge the valuable discussions and feedback from the SE$^2$A B4.2. group, particularly Susanna Baars and Ihar Antonau.
This work was conducted using Protégé.

\newpage

\section*{Biographies}

\begin{wrapfigure}{l}{0.25\textwidth}
    \includegraphics[width=0.9\linewidth]{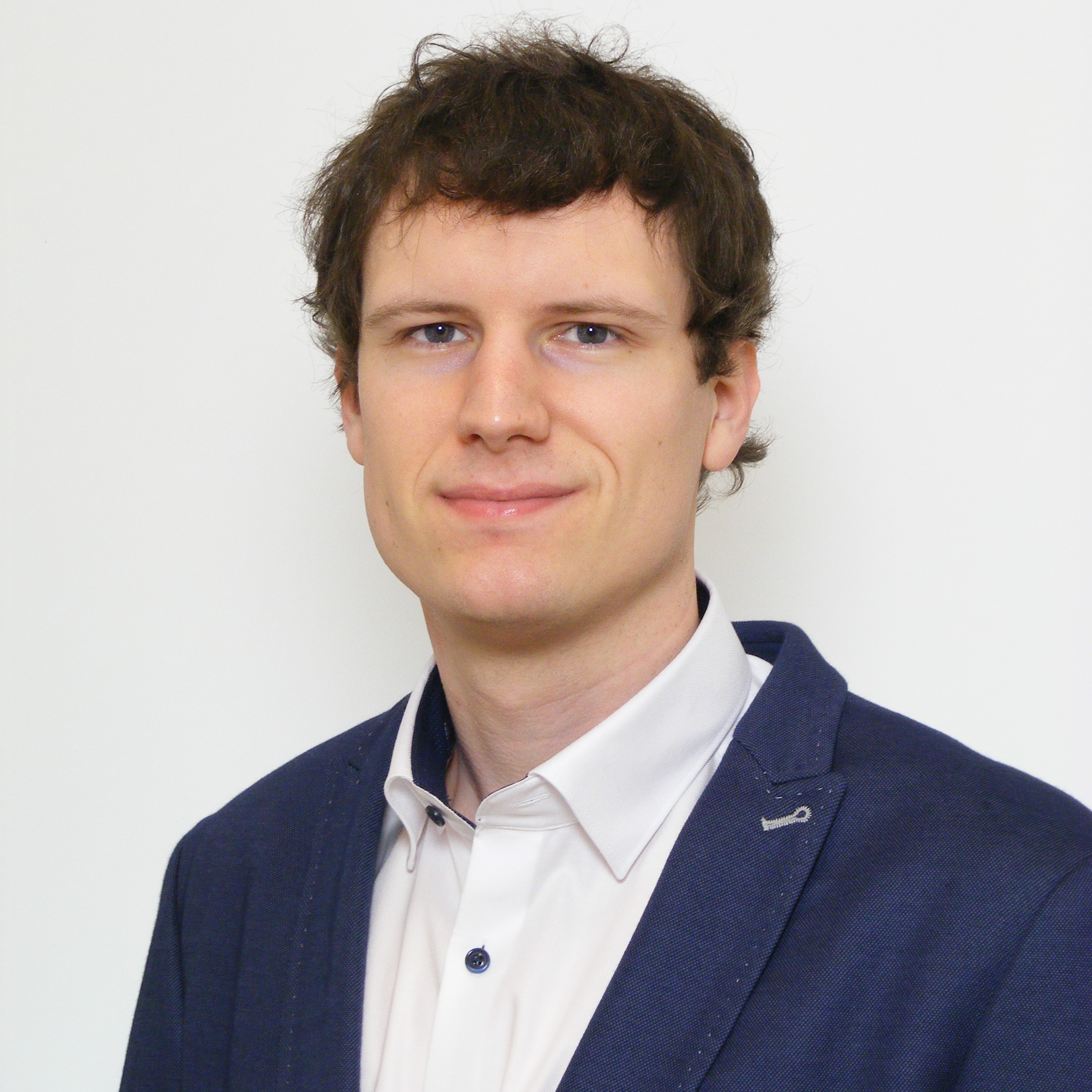} 
    \caption{Tim Wittenborg}
    \label{fig:wrapfig}
\end{wrapfigure}

\paragraph{Tim Wittenborg} is a PhD Student at the L3S Research Center, working in the SE²A - Sustainable and Energy-Efficient Aviation Cluster of Excellence and on the Open Research Knowledge Graph.
He has a background in Mechanical Engineering and Computer Science, completing his Master degree at the University of Paderborn in 2022 and spending 10 months as a mechanical engineering PhD student in the information management group of the product creation department.
Since progressing to Leibniz University Hannover in January 2023, he studies under professor Sören Auer, focussed on developing knowledge infrastructure and communities.
He has published a paper on such infrastructure supporting the systematic literature process and founded a non-profit organization to support the establishment of a knowledge community for distributed scientific communication.

\wrapfill

\begin{wrapfigure}{r}{0.25\textwidth}
    \includegraphics[width=0.9\linewidth]{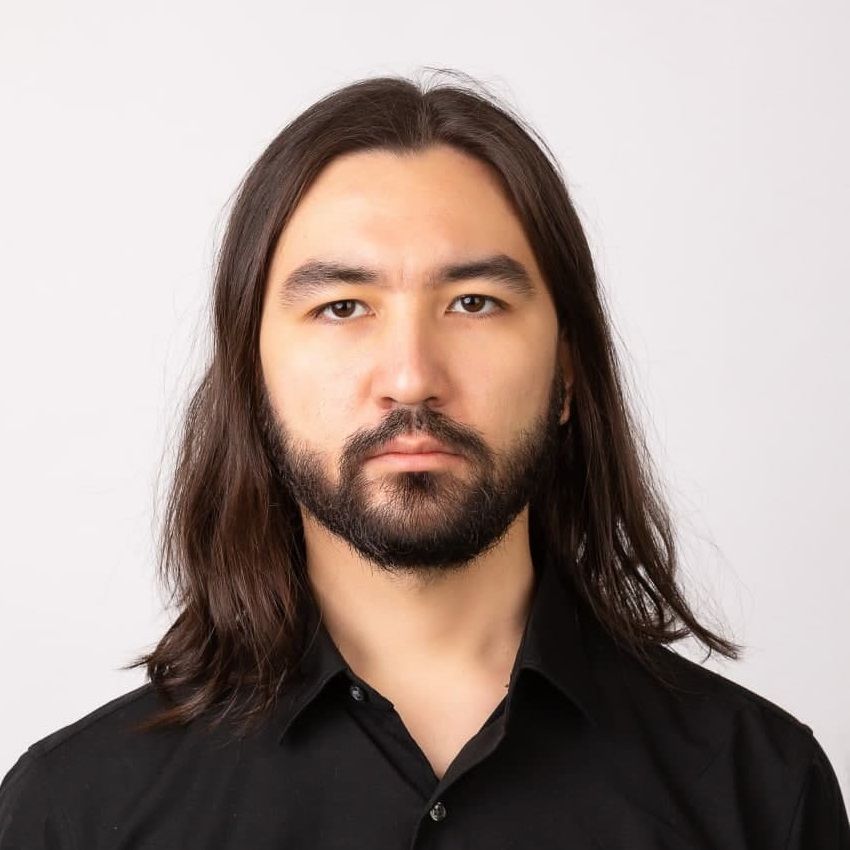} 
    \caption{Ildar Baimuratov}
    \label{fig:Ildar}
\end{wrapfigure}

\paragraph{Ildar Baimuratov} earned a Master’s degree in Logic from St. Petersburg University in 2014, followed by a PhD in Computer Science from ITMO University in 2020. His doctoral thesis was focused on clustering hyperparameter tuning, while simultaneously working on projects involving machine learning and knowledge engineering. Since 2022, Ildar has been a postdoctoral researcher at the joint lab of the L3S Research Center and TIB - Leibniz Information Centre for Science and Technology, as well as the Sustainable and Energy-Efficient Aviation (SE$^{2}$A) Cluster of Excellence. His research emphasizes advancing knowledge extraction, representation, and utilization in scholarly contexts — such as automating peer review — and engineering domains, including building construction and aerospace engineering.

\wrapfill

\begin{wrapfigure}{l}{0.25\textwidth}
    \includegraphics[width=0.9\linewidth]{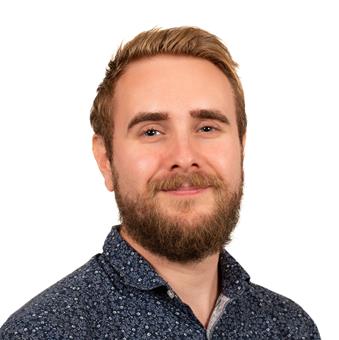} 
    \caption{Ludvig Knöös Franzén}
    \label{fig:Ludvig}
\end{wrapfigure}

\paragraph{Ludvig Knöös Franzén} is an assistant professor in aircraft design at the division of Fluid- and Mechatronic Systems in the department of management and engineering at Linköping University (LiU) in Sweden. He holds a Bachelor’s degree in Mechanical Engineering, as well as a Master's degree in Aeronautical Engineering from LiU. In 2023, Ludvig also defended his Ph.D. thesis at LiU with the title “A System of Systems View in Early Product Development – An Ontology-based Approach”. Ludvig's research is primarily focused on aircraft design and design space explorations from a System-of-Systems perspective. He is currently a work package leader in the COLOSSUS (Collaborative System of Systems Exploration of Aviation Products, Services \& Business Models) project funded by the European Union Horizon Europe program under grant agreement
No. 101097120. Furthermore, Ludvig received a best Ph.D. project award in 2020 on the Doctoral Consortium on Knowledge Discovery, Knowledge Engineering and Knowledge Management (DC3K/IC3K).

\wrapfill

\begin{wrapfigure}{r}{0.25\textwidth}
    \includegraphics[width=0.9\linewidth]{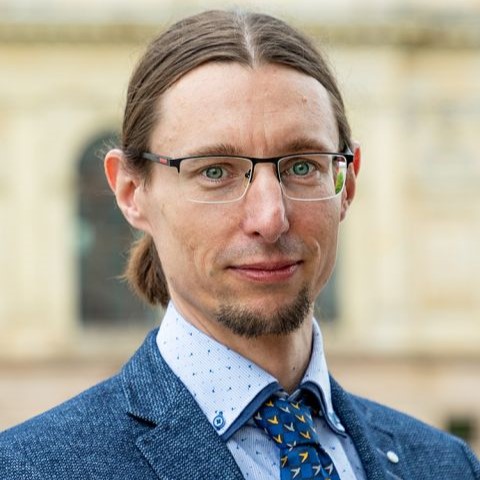} 
    \caption{Ingo Staack}
    \label{fig:Ingo}
\end{wrapfigure}

\paragraph{Ingo Staack} is professor for aircraft design at the Technische Universität Braunschweig. With a background in Mechanical Engineering at the \gls{tu} Munich. He obtained his PhD in Aerospace Engineering from Linköping University, Sweden in 2016, His research encompasses a wide range of engineering design topics, ranging from early conceptual design to protoptyping of flying demonstrators, extending aircraft design through systems engineering (SE) and system-of-systems engineering (SoSE) concepts for innovative and complex product development. His institute is a member of the Aeronautics Research Centre Niedersachsen (NFL) and the Sustainable and Energy-Efficient Aviation (SE$^{2}$A) Cluster of Excellence. 

\wrapfill

\begin{wrapfigure}{l}{0.25\textwidth}
    \includegraphics[width=0.9\linewidth]{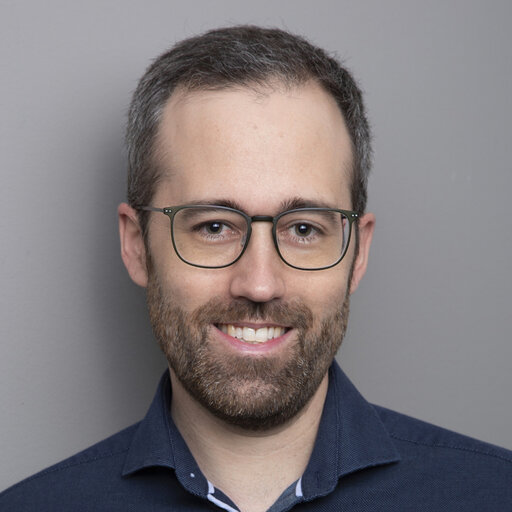} 
    \caption{Ulrich Römer}
    \label{fig:Ulrich}
\end{wrapfigure}

\paragraph{Ulrich Römer} is associate professor for Modeling of Complex Systems under Uncertainty in Mobility at TU Braunschweig, Germany. He holds a double degree in Electrical Engineering and Information Technology from TU Darmstadt, Germany, and General Engineering from École Centrale de Lyon, France. His PhD, completed in 2015 at TU Darmstadt, focused on uncertainty quantification in computational magnetics. He is a member of the Aeronautics Research Centre Niedersachsen and founding PI as well as board member of the Sustainable and Energy-Efficient Aviation (SE$^{2}$A) Cluster of Excellence. His research focuses on computational and data-driven modeling and design, with an emphasis on uncertainty quantification. 

\wrapfill

\begin{wrapfigure}{r}{0.25\textwidth}
    \includegraphics[width=0.9\linewidth]{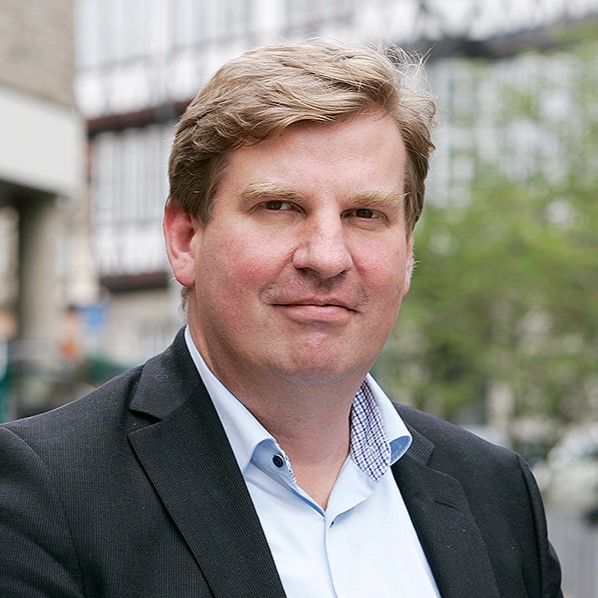} 
    \caption{Sören Auer}
    \label{fig:Sören}
\end{wrapfigure}

\paragraph{Sören Auer} is Director of TIB - Leibniz Information Centre for Science and Technology and professor for Data Science and Digital Libraries at Leibniz University of Hannover.
Prof. Auer has made important contributions to semantic technologies, knowledge engineering, and information systems. 
He is the author (resp. co-author) of over 200 peer-reviewed scientific publications. 
He has received several awards, including an ERC Consolidator Grant from the European Research Council, a SWSA ten-year award, and the ESWC 7-year Best Paper Award. 
He is a co-founder of high-potential research and community projects such as the Wikipedia semantification project DBpedia, the Open Research Knowledge Graph ORKG.org, and the innovative technology start-up eccenca.com.

\wrapfill

\bibliography{main}

\end{document}